\documentclass[a4paper,11pt]{article}
\usepackage{jinstpub} 

\usepackage{amsmath}
\usepackage{subfigure}
\usepackage{booktabs}
\usepackage{multirow}
\usepackage{textcomp}
\usepackage{siunitx}
\usepackage{orcidlink}
\usepackage{hyperref}

\title{\boldmath Quality Control for the LGAD sensors of the ATLAS High Granularity Timing Detector}

\author[1,a,*]{K.~Ma\orcidlink{0009-0004-7076-0889}\note[\relax$^{*}$]{Corresponding author.}\note[\relax$^{a}$]{These authors contributed equally to this work.},}
\emailAdd{kuo.ma@cern.ch}
\author[2,3,a]{W.~Sun\orcidlink{0000-0003-4002-0199},}
\author[4]{M.~Ait~Tamlihat\orcidlink{0000-0002-1322-4666},}
\author[5,6]{H.M.~Alfanda,}
\author[7]{N.~Atanov\orcidlink{0000-0003-1744-5439},}
\author[7]{O.~Atanova\orcidlink{0009-0009-5280-3095},}
\author[8]{I.~Azzouzi,}
\author[2]{J.~Barreiro~Guimar\~{a}es~Da~Costa\orcidlink{0000-0003-2387-0386},}
\author[9]{L.T.~Bazzano~Hurrell\orcidlink{0000-0002-8985-6934},}
\author[10]{T.~Beau\orcidlink{0000-0002-2022-2140},}
\author[11]{D.~Benchekroun\orcidlink{0000-0001-5196-8327},}
% ...
\affiliation[1]{Department of Modern Physics and State Key Laboratory of Particle Detection and Electronics, University of Science and Technology of China, Hefei, China}
\affiliation[2]{Institute of High Energy Physics, Chinese Academy of Sciences, Beijing, China}
\affiliation[3]{University of Chinese Academy of Sciences (UCAS), Beijing, China}
\affiliation[4]{Facult\'e des sciences, Universit\'e Mohammed V, Rabat, Morocco}
\affiliation[5]{State Key Laboratory of Dark Matter Physics, Tsung-Dao Lee Institute, Shanghai Jiao Tong University, Shanghai, China}
\affiliation[6]{State Key Laboratory of Dark Matter Physics, Key Laboratory for Particle Astrophysics and Cosmology (MOE), Shanghai Key Laboratory for Particle Physics and Cosmology, School of Physics and Astronomy, Shanghai Jiao Tong University, Shanghai, China}
\affiliation[7]{Affiliated with an international laboratory covered by a cooperation agreement with CERN}
\affiliation[8]{Moroccan Foundation for Advanced Science Innovation and Research (MAScIR), Rabat, Morocco}
\affiliation[9]{Institut de F\'isica d'Altes Energies (IFAE), Barcelona Institute of Science and Technology, Barcelona, Spain}
\affiliation[10]{LPNHE, Sorbonne Universit\'e, Universit\'e Paris Cit\'e, CNRS/IN2P3, Paris, France}
\affiliation[11]{Facult\'e des Sciences Ain Chock, Universit\'e Hassan II de Casablanca, Casablanca, Morocco}
% ...

\abstract{The High Granularity Timing Detector (HGTD) currently in the production phase will mainly mitigate the pile-up effect in the ATLAS forward region and serve for bunch-by-bunch luminosity measurements in the high-luminosity (HL) phase of the Large Hadron Collider (LHC). The HGTD utilizes Low Gain Avalanche Detector (LGAD) technology, covering the pseudorapidity range between 2.4 and 4.0, and provides high-precision timing information to distinguish collisions occurring close in space but well-separated in time. A pre-production batch of 8-inch silicon wafers was fabricated and a systematic quality control test campaign was carried out. This paper describes the Quality Control (QC) required for full-size (15 $\times$ 15 array) sensors, including leakage-current and breakdown measurements performed on all sensors at the vendor sites, as well as more detailed electrical characterizations at the participating ATLAS-HGTD institutes. In addition, the electrical characterization of full-size sensors after irradiation is reported.}


\begin{document}
\maketitle
\flushbottom

\section{Introduction}
\label{sec1}

The High Luminosity phase of the Large Hadron Collider (HL-LHC) will reach an instantaneous luminosity of 7.5~$\times$~10$^{34}$ cm$^{-2}$s$^{-1}$~\cite{CERN-LHCC-2020-007}. A new High Granularity Timing Detector (HGTD) will be installed in front of the end-cap and forward calorimeter at 2.4~$<$~$|\eta|$~$<$~4.0 to measure charged-particle trajectories in time and space, mitigating pile-up and increasing the reach of the new all-silicon Inner Tracker (ITk)~\cite{ATLAS:2017svb}. The Low Gain Avalanche Detector (LGAD)~\cite{SADROZINSKI201618}, based on an n-in-p structure with a p-type gain layer beneath the np junction on the top side and reversely biased from the back side, was chosen as the sensing technology. The internal gain degrades after irradiation due to acceptor removal in the LGAD gain layer~\cite{Kramberger2015, FERRERO201916}. In order to partially compensate for the loss of gain, the operating bias voltage should be increased but is limited to 550~V (corresponding to an average electric field of \SI{11}{V/\micro m} for an active sensor thickness of \SI{50}{\micro m}), to avoid destructive single-event burnout (SEB) from localized high ionization charges~\cite{SEB_2022, Beresford_2023}. A single bias voltage will be applied to the back side of the two LGAD sensors assembled in one module with two readout ASIC chips during the operation. The sensors are required to have a time resolution better than 40 ps at the start-of-life for a Minimum Ionizing Particle (MIP). After receiving an end-of-life fluence of 2.5~$\times$~$\mathrm{10^{15}}$~$\mathrm{n_{eq}/cm^2}$ (1 MeV neutron equivalent fluence), the sensors should maintain a time resolution better than 50~ps, and a most probable collected charge of at least 4~fC.

During the past decade, LGAD technology has been widely developed by many producers, including CNM~\cite{PELLEGRINI201412}, FBK~\cite{SOLA2019360}, HPK~\cite{GALLOWAY201919}, BNL~\cite{GIACOMINI201952}, IHEP-IME~\cite{ZHAO2022166604} and USTC-IME~\cite{LI2022167008}. The ATLAS-HGTD collaboration conducted a systematic evaluation of the most promising LGAD options from multiple vendors, covering both laboratory measurements and test-beam results~\cite{MA2024169237}. Ultimately, two sensor designs were adopted: the IHEP-IME version (designed by IHEP and fabricated by IME) and the USTC-IME version (designed by USTC and fabricated by IME). The two designs are independent, although they share the same production facility at IME. Taking into account an estimated assembly yield of 75$\%$, ATLAS-HGTD will require approximately 22000 full-size (15~$\times$~15 array) good sensors. 

To validate the feasibility of final production, a pre-production batch was produced between early 2023 and the end of 2024, corresponding to at least 5\% of the total quantity of sensors. The pre-production was accompanied by systematic quality control tests. One area of quality control is the electrical characterization of 15~$\times$~15 array sensors. Company Quality Control (CQC) comprises Current-Voltage ($I-V$) characterization at the vendor sites, using a single probe needle or a 15~$\times$~1 probe card. Another aspect of quality control, called Sensor Quality Control (SQC) on a subset of sensors after Under Bump Metallization (UBM) deposition and dicing, includes per-pad $I-V$ tests at the participating ATLAS-HGTD institutes using probe needles under conditions identical to CQC and a matrix test using a 15~$\times$~15 probe card. The gain layer depletion voltage is extracted from the Capacitance-Voltage ($C-V$) data. The results of the SQC measurements are used to validate and refine the CQC measurement setups and the CQC procedures to be performed on all sensors during the production phase. Electrical characterization of sensors after irradiation is performed, to validate the radiation hardness and complement dedicated irradiation tests on QC-TS reported previously~\cite{VELKOVSKA2026171195}. This paper is organized as follows: Section~\ref{Acceptance criteria} lists the acceptance criteria, Section~\ref{Samples and experimental set-up} describes the samples and experimental setup, Section~\ref{CQC} presents the results of the CQC and Section~\ref{SQC} shows the results of the SQC. The results of irradiated sensors are discussed in Section~\ref{Irradiation tests}. Section~\ref{Conclusion} gives the conclusions.

\section{Observable definitions and acceptance criteria}
\label{Acceptance criteria}

In order for sensors to be accepted during the CQC, they must meet the following conditions at \SI{20}{\celsius}: 

\begin{itemize}
    \item[$\bullet$]
    Pad breakdown voltage: The breakdown voltage of a single pad ($V_\mathrm{bd,~pad}$) is defined as the bias voltage at which the leakage current ($I_\mathrm{bd,~pad}$) reaches 500~nA. The corresponding current threshold for simultaneously measuring 15 pads (15~$\times$~1 array) is \SI{7.5}{\micro A}. In order to obtain $V_\mathrm{bd,~pad}$, two adjacent voltage points with current below and above 500 nA are used for linear interpolation. The $V_\mathrm{bd,~pad}$ is required to exceed the full depletion voltage of the pad by $D~\times$ \SI{2}{\volt/\micro\meter}, where $D$ is the thickness of the sensitive region, i.e, \SI{50}{\micro m}. The excessive electric field of \SI{2}{V/\micro m} is to ensure approaching saturation velocity of carriers. The RMS($V_\mathrm{bd,~pad}$)/AVERAGE($V_\mathrm{bd,~pad}$) over a 15~$\times$~15 array sensor is required to be less than 0.05, where RMS($V_\mathrm{bd,~pad}$) is calculated as the standard deviation of $V_\mathrm{bd,~pad}$ and AVERAGE($V_\mathrm{bd,~pad}$) is defined as the mean $V_\mathrm{bd,~pad}$ of the sensor.
    \item[$\bullet$] 
    Sensor breakdown voltage: The sensor breakdown voltage ($V_\mathrm{bd}$) is defined by the mean $V_\mathrm{bd,~pad}$ of all 225 pads for USTC-IME sensors and the mean $V_\mathrm{bd,~pad}$ the 15 rows measured by the $15\times 1$ probe card for IHEP-IME sensors. For USTC‑IME sensors and IHEP‑IME sensors, the maximum $V_\mathrm{bd}$ difference between the whole pre-production sensors from the same vendor must not exceed 8\% of the average $V_\mathrm{bd}$ of that vendor's sensors. These requirements result in acceptance criteria of 165~V $<$ $V_\mathrm{bd}$ $<$ 195~V for IHEP-IME sensors, and of 174~V $<$ $V_\mathrm{bd}$ $<$ 205~V for USTC-IME sensors. For the final production, the requirement was later unified to 165~V $<$ $V_\mathrm{bd}$ $<$ 195~V, where $V_\mathrm{bd}$ is redefined from the mean to the minimum of the per pad or per row breakdown voltages, in order to reject sensors with early breakdown. Tested using the pre-production data, this modification rejects about 5\% more of the sensors for batches with nominal gain layer implantation dose.  
    \item[$\bullet$]
    Current spread: For per-pad tests, Max($I$ at 0.8~$\cdot~V_\mathrm{bd}$)/Min($I$ at 0.8~$\cdot~V_\mathrm{bd}$) $\leq$ 3, where Max($I$ at 0.8~$\cdot~V_\mathrm{bd}$) or Min($I$ at 0.8~$\cdot~V_\mathrm{bd}$) corresponds to the maximum or minimum current when the bias voltage of the sensor is equal to 0.8$\cdot$$V_\mathrm{bd}$. For the single-needle measurement the pads in the outermost two rings are excluded from the consideration because it is not possible to ground the guard ring. For the $15\times 1$ probe-card tests:  Max($I_\mathrm{{15pad}}$ at $0.8\cdot V_{\mathrm{bd}}$)/Min($I_\mathrm{{15pad}}$ at $0.8\cdot V_{\mathrm{bd}}$) $\leq$ 2.5. This tightened cut accounts for the fact that individual pad-current outliers can be concealed in the combined leakage of the 15 pads in a column (see also section~\ref{CQC15x1results}).
\end{itemize}

As the USTC-IME sensors are delivered on wafers that need to be further post-processed (UBM and dicing), they need to fulfill the additional requirement of containing at least 18 sensors meeting the specifications above. 

During the SQC, two parameters are defined to quantify differences in the per-pad tests: the difference of $V\mathrm{_{bd,~pad}}$, calculated as $V\mathrm{_{bd,~pad}}$ measured by the CQC minus $V\mathrm{_{bd,~pad}}$ measured by the SQC; the ratio of $I$ at 0.8~$\cdot~V_\mathrm{bd,~pad}$, calculated as the ratio of $I$ at 0.8~$\cdot~V_\mathrm{bd,~pad}$ measured by the CQC to $I$ at 0.8~$\cdot~V_\mathrm{bd,~pad}$ measured by the SQC.

The gain-layer depletion voltage ($V_\textrm{gl,~pad}$) is evaluated during the SQC by $C-V$ measurements~\cite{Ferrero:2021lwf}. For matrix tests of the main sensors, the relative spread of $V_\mathrm{gl,~pad}$, defined in the same manner as the the spread of breakdown voltage, i.e, the ratio of the standard deviation of $V_\mathrm{gl,~pad}$ (RMS($V_\mathrm{gl,~pad}$)) to the mean $V_\mathrm{gl,~pad}$ of the sensor (AVERAGE($V_\mathrm{gl,~pad}$)). The requirement is required to be smaller than 0.005. During production the $V_\textrm{gl,~pad}$ acceptance test will be performed by $C-V$ measurements on $\ge$ 20 single LGAD sensors (from the test structures) per wafer. The standard deviation of the $V_\textrm{gl,~pad}$ is required to be within 2$\%$. As part of SQC, irradiated test structures and full-size sensors are tested to assess the radiation hardness of the gain layer and the isolation structures between pads. It is required that after the fluence of 2.5~$\times$~$\mathrm{10^{15}}$~$\mathrm{n_{eq}/cm^2}$, the gain with bias voltage of 100 V should be greater than 2.7, and the $V_\textrm{gl,~pad}$ should be above 16.7 V, as evaluated at \SI{20}{\celsius}; the inter-pad resistance should be greater than \SI{1}{\mega\ohm} at the operation bias voltage, as evaluated at \SI{-30}{\celsius}. During the prototyping phase, it has been demonstrated that the leakage current of \SI{1}{\micro A} corresponds to a charge collection of 5-7 fC. Based on the $I-V$ results of the irradiated main sensors, $V$ at \SI{1}{\micro A} (defined as the bias voltage where the pad's leakage current is equal to \SI{1}{\micro A}) was extracted, which shall be smaller \SI{550}{\volt} to avoid the SEB effect.

\section{Samples and experimental set-up}
\label{Samples and experimental set-up}

\subsection{Samples}

\begin{figure}[htbp]
  \centering
  \subfigure[]{
    \includegraphics[width=0.28\textwidth]{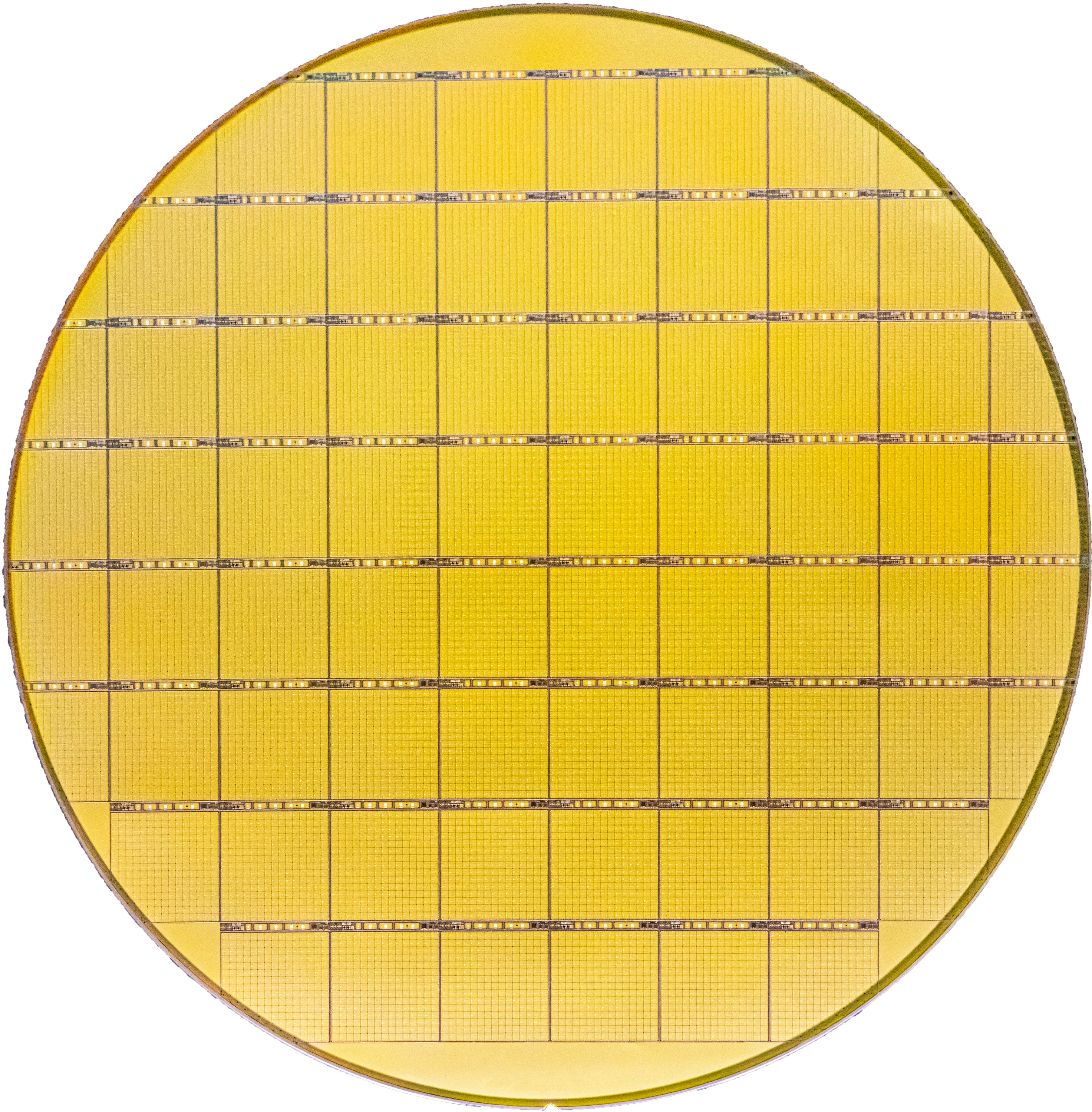}}
    \subfigure[]{
	\includegraphics[width=0.40\textwidth]{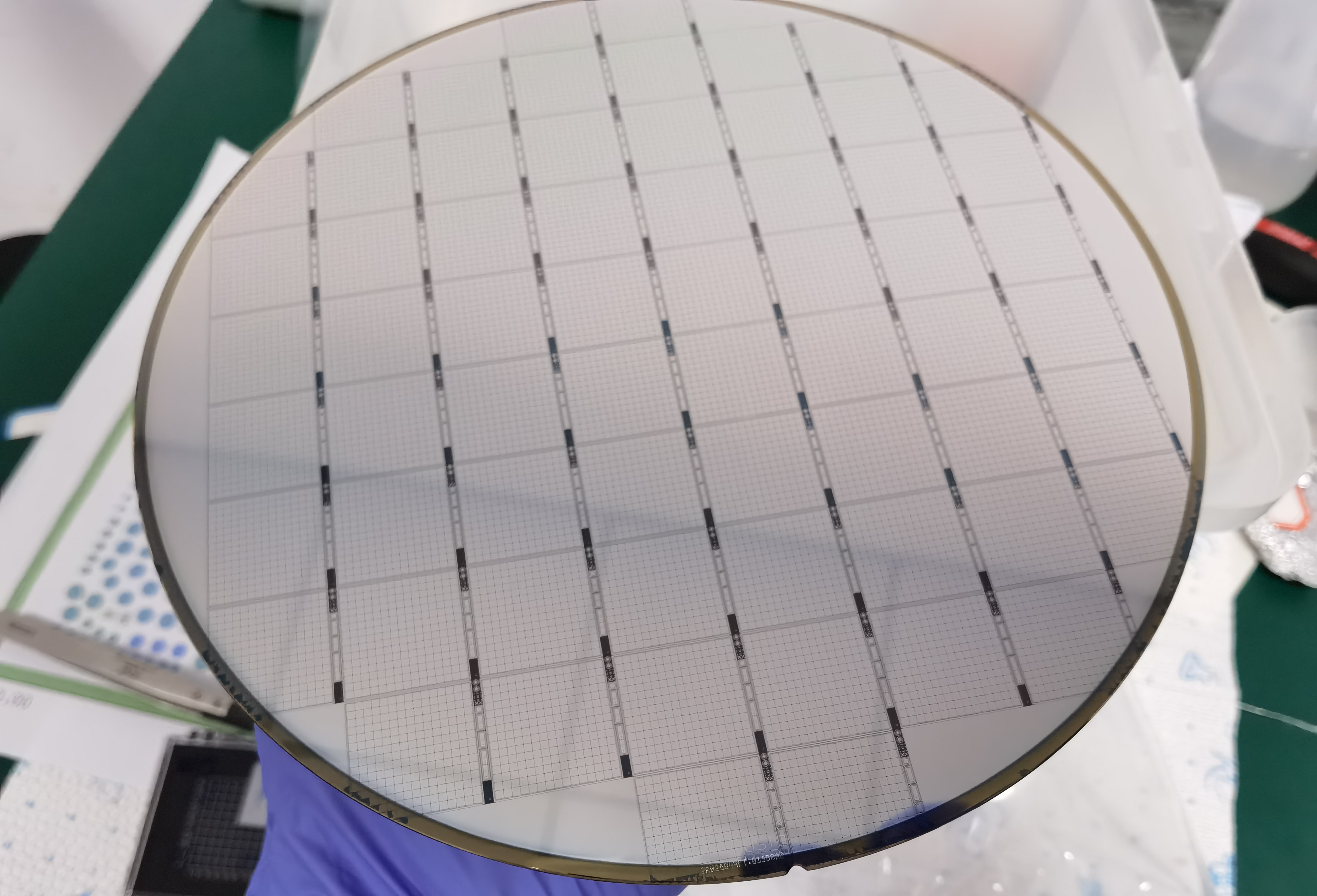}}
    \subfigure[]{
    \includegraphics[width=0.25\textwidth]{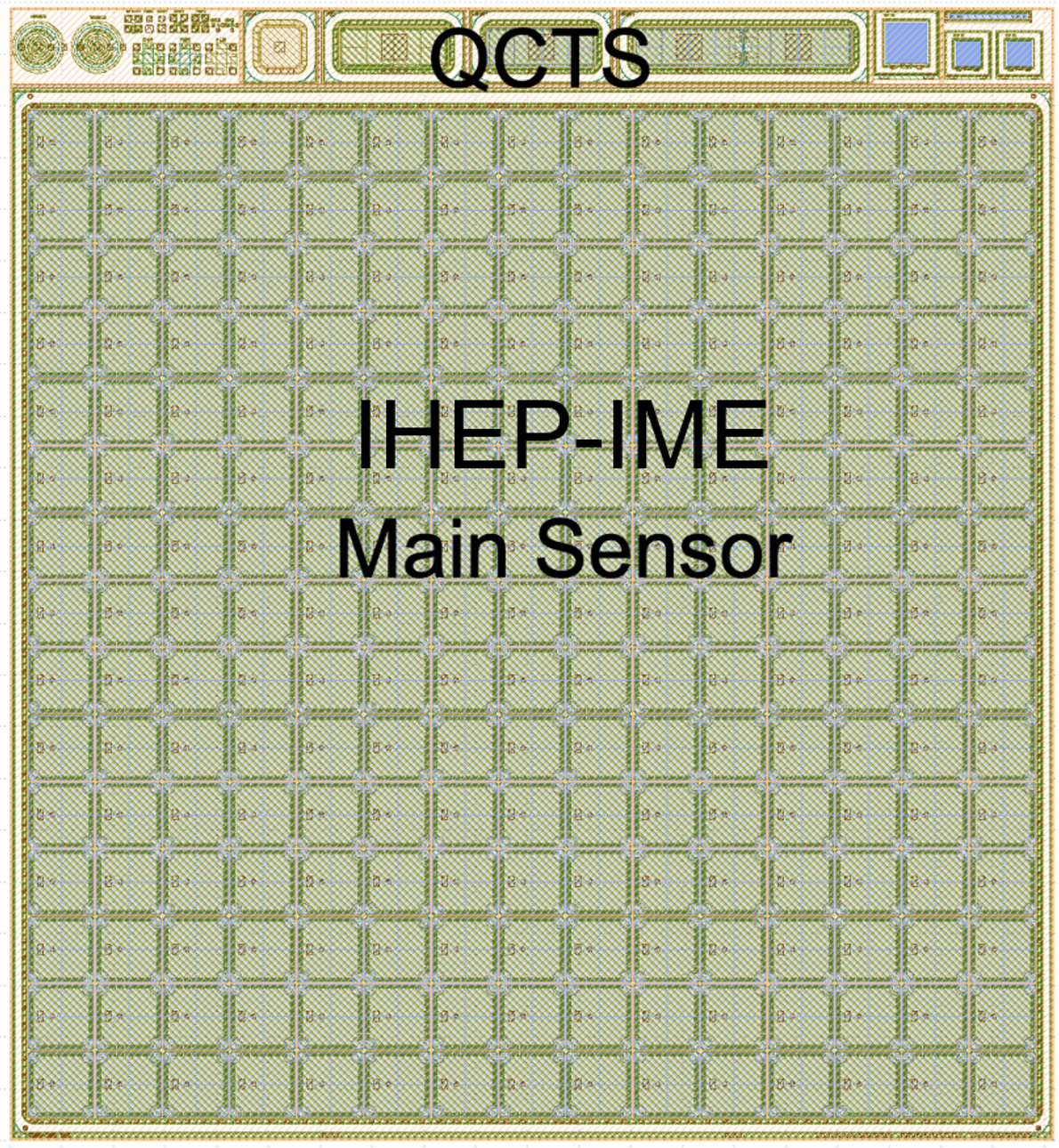}}
  \caption{(a) Photograph of an IHEP-IME wafer. (b) Photograph of a USTC-IME wafer. (c) A reticle includes a main sensor and a QC-TS of IHEP-IME sensor (USTC-IME sensor has the same top layer layout, except that the QC-TS is placed vertically).}
  \label{fig:HGTDPreWafer}
\end{figure}

The sensors were fabricated on 8-inch epitaxial silicon wafers, with an active thickness of \SI{50}{\micro m} and a total thickness of \SI{775}{\micro m}. The IHEP-IME and USTC-IME wafers share a similar layout: each wafer contains 52 complete reticles, consisting of a main sensor and a Quality Control Test Structure (QC-TS), as shown in Fig. \ref{fig:HGTDPreWafer}. Each main sensor consists of 15~$\times$~15 pads with a pad size of \SI{1.3}{mm}~$\times$~\SI{1.3}{mm} separated by a gap (without gain layer) of \SI{50}{\micro m}. Each QC-TS contains a single LGAD, a 1~$\times$~2 LGAD array, a PIN diode, and other structures for the control of different process parameters. 

During the pre-production phase, 90 IHEP-IME wafers were fabricated in several batches. The process parameters, including detector thickness and surface processing of later batches, were iteratively adjusted according to the QC results obtained from earlier batches. One wafer was selected as the final design and an additional of 22 wafers were produced for the pre-production. The results presented in this paper for IHEP-IME pre-production are based on these 23 wafers. For USTC-IME design, a total of 9 wafers were fabricated for pre-production.

\subsection{Experimental setup}
\label{setup}
An overview of the sensor CQC and SQC measurement setup is presented in Table~\ref{table:probestations} and summarized below. 
\begin{table}[ht]
\begin{tabular}{lcccc}
\toprule
   & IHEP CQC & IHEP SQC & USTC CQC & USTC SQC \\
\midrule
\multirow{2}{*}{Probe station} & MPI & Cascade  & Cascade  & Apollowave  \\
 &  TS2000-HP  & summit 200 & summit 200  
& alpha-200CS  \\
\hline
Chuck temperature & \SI{22}{\celsius} & \SI{22}{\celsius} & Not monitored & \SI{20}{\celsius} \\
\hline
\multirow{2}{*}{SMU device} & Keysight  & Keithley & Keysight  & Keithley \\
& B1505 &  2665A &  B1500 & 2410 and 6482 \\
\hline
Voltage range & 0 -- \SI{230}{V} & 0 -- \SI{240}{V} & 0 -- \SI{230}{V} & 0 -- \SI{200}{V} \\
\hline
Voltage step & 2 V & 5 V & 2 V & 5 V \\
\hline
Settling time & 0.03 s & 0.1 s & 0.03 s & 0.2 s \\
\hline
Compliance current & \SI{50}{\micro A} &  \SI{10}{\micro A} & \SI{60}{\micro A} & \SI{600}{\micro A} \\
\hline
Gurad ring  & grounded & floating & floating & floating \\
\hline
Probe & 15~$\times$~1 card, 1 needle & 1 needle & 3 needles & 15~$\times$~15 card \\
\bottomrule
\end{tabular}
\caption{Settings of the 4 measurement systems for the CQC and SQC.}
\label{table:probestations}
\end{table}
The CQC of the IHEP-IME wafers consisted of full wafer level testing at IME, followed by data analysis at IHEP, while the CQC of the  USTC-IME wafers was carried out at USTC at full wafer level. 

Initially,  the per-pad testing was exercised for IHEP CQC. With voltage steps of 5~V and a leakage current compliance of \SI{10}{\micro A}, the average test time per sensor is about 100 minutes. To speed-up the test, a 15~$\times$~1 probe card was used to measure the sum of the leakage currents of one column (15 pads). The voltage step is 2~V for the 15~$\times$~1 probe card test. 
The chuck temperature was not controlled or monitored for the USTC CQC setup. The measurement is done in a lab with AC (air temperature about \SI{25}{\celsius} and relative humidity \(<50\%\)). Three pads were connected to three individual channels of the B1500 analyzer through probe needles for simultaneous measurements, so the chuck moves 75 times, instead of 225 times, to test one sensor.
The guard rings (GRs) were tested individually by a single needle after finishing all pad measurements. In total, the average measurement time for one sensor is about 5 minutes. A new probe station with temperature-controlled chuck has been installed and commissioned for the final production CQC at USTC.

The 15~$\times$~15 matrix test setup at USTC is based on a manual probe station with a thermal chuck as described in detail elsewhere~\cite{GE2021165400}. All measurements of the unirradiated sensors were conducted at \SI{20}{\celsius} and a relative humidity below 38$\%$. According to the testing objectives of the irradiated sensors, the chuck was kept at \SI{20}{\celsius} or \SI{-30}{\celsius}. 
The capacitance was tested by an LCR meter (Agilent E4980 A) with a high-voltage decoupling circuit. 

\begin{figure}[htbp]
	\centering
        \subfigure[]{\includegraphics[width=0.4\textwidth]{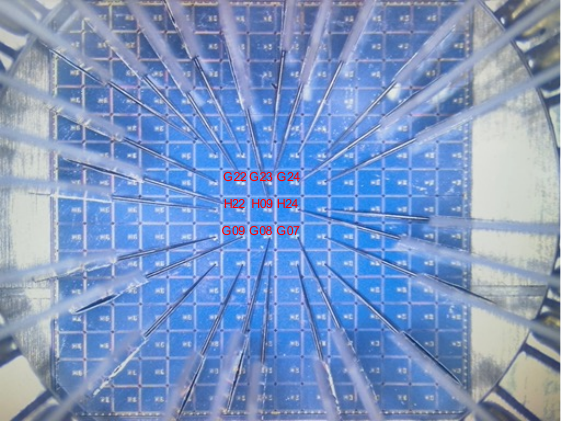}}
        \subfigure[]{\includegraphics[width=0.4\textwidth]{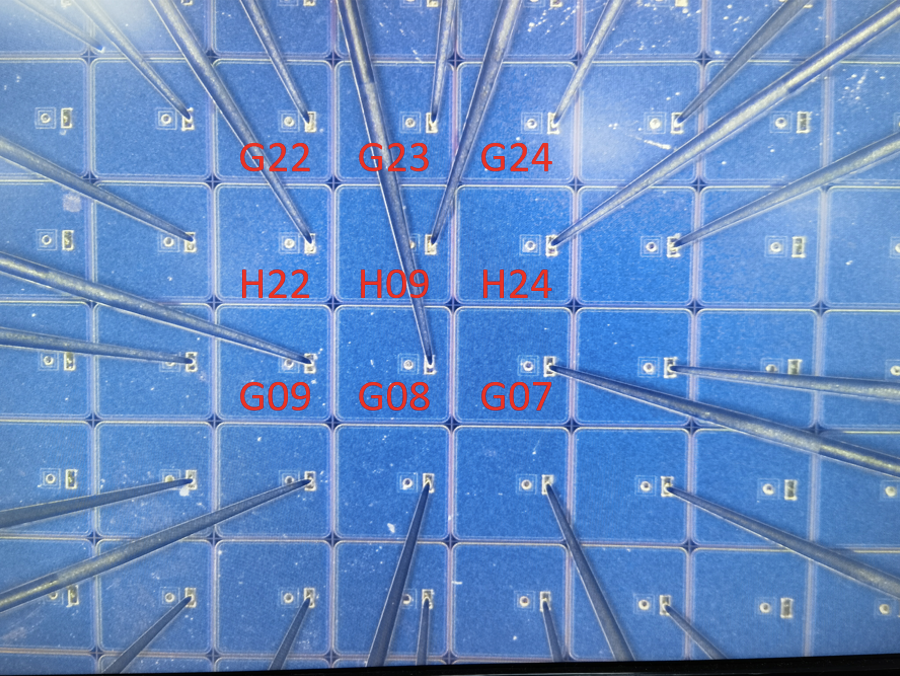}}
        \caption{The micrograph of a 15~$\times$~15 sensor connected with 5~$\times$~5 probe card. (a) The whole sensor. (b) The center region of the sensor. The central pad is labeled as H09, while the eight adjacent pads are labeled as G07, G08, G09, G22, G23, G24, H22 and H24.}
        \label{USTCIPsetup}
\end{figure}

The inter-pad resistance measurement of the irradiated sensors was tested using the 5~$\times$~5 probe card at \SI{-30}{\celsius}. Fig.~\ref{USTCIPsetup} shows the 25 pads in the middle of the sensor connected to a 5~$\times$~5 probe card. The central pad (H09) was biased at 0 or 1 V in addition using a Keithley 6482 channel, and the leakage current of the eight adjacent pads (G07, G08, G09, G22, G23, G24, H22 and H24) was measured using another Keithley 6482 channel. The outermost 16 pads served as a guard ring (GR)/shield.

\subsection{Temperature dependence of breakdown voltage }
\label{SVBDvsT}

\begin{figure}[htbp]
	\centering
	\subfigure[]{
	\includegraphics[width=0.45\textwidth]{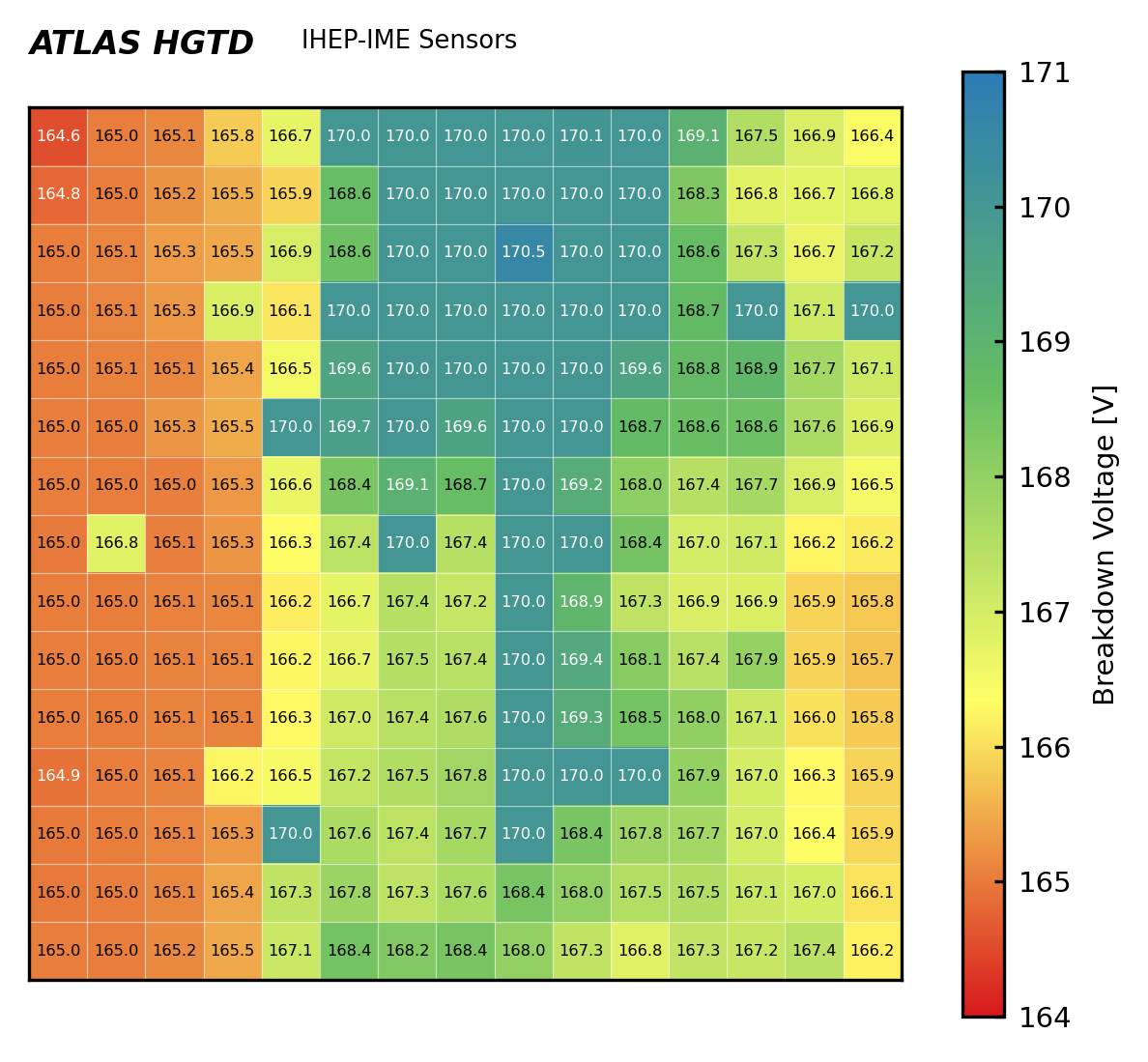}}
    \subfigure[]{
	\includegraphics[width=0.5\textwidth]{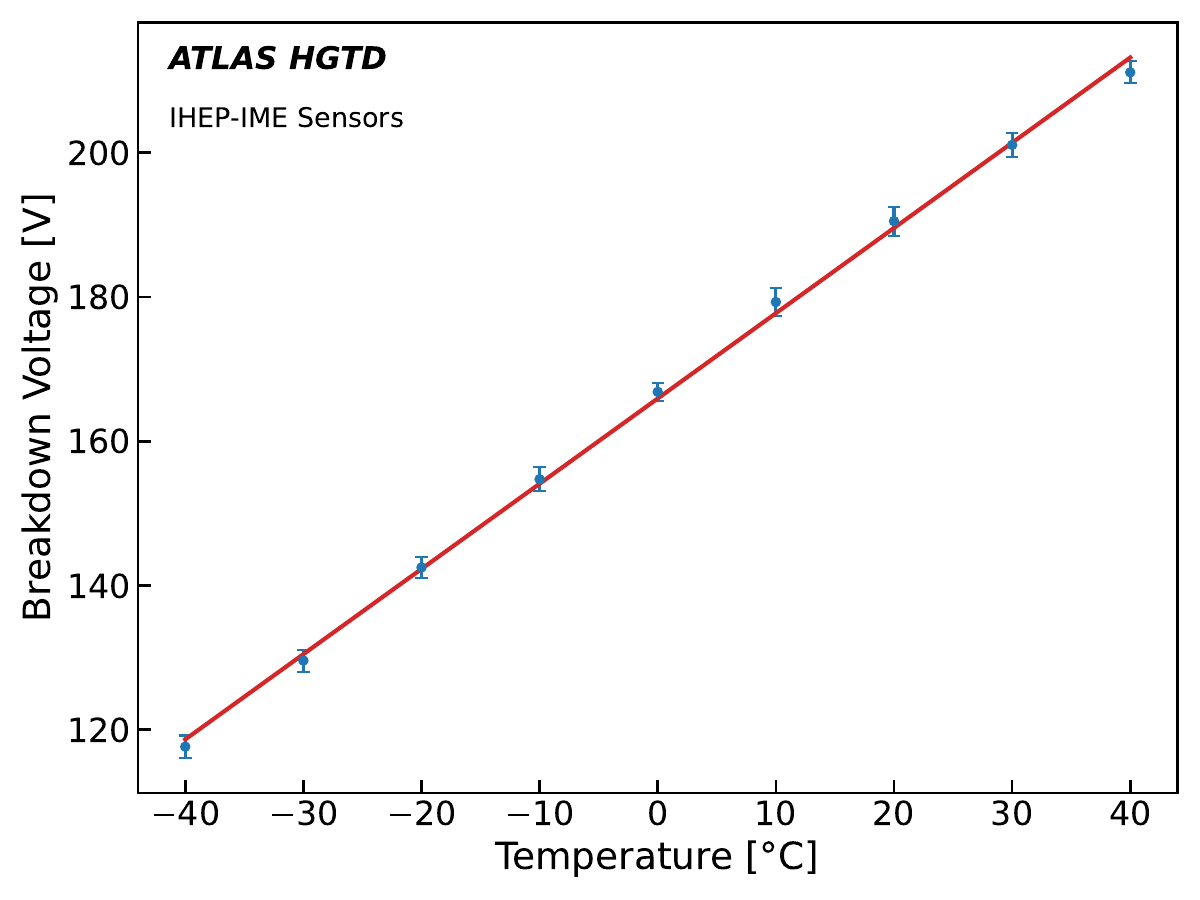}\label{VBDvsT}}
    \caption{Temperature dependence of the IHEP-IME breakdown voltage.(a) Two-dimensional distribution of $V_{\mathrm{bd,pad}}$ at \SI{0}{\celsius}. (b)  $V_{\mathrm{bd,pad}}$ distribution as a function of temperature.
    }
    \label{fig:IHEPVT}
\end{figure}

To investigate the temperature dependence of the breakdown voltage, dedicated $I-V$ scans were performed on the diced sensors over a temperature range of \SIrange{-40}{40}{\celsius} in steps of \SI{10}{\celsius} with the IHEP SQC setup and IHEP-IME sensors. The same setup of the per-pad test and the same definition of the breakdown voltage as previously described were used throughout the measurements. The distributions of $V_\mathrm{bd,~pad}$ at \SI{0}{\celsius} for a representative sensor, together with the breakdown voltages measured at all temperatures, are shown in Fig.~\ref{fig:IHEPVT}.

At each temperature, the distribution of
$V_{\mathrm{bd,pad}}$ over the 225 pads was fitted with a Gaussian function. The fitted Gaussian mean is used as the representative breakdown voltage at that temperature. The vertical error bar is
the fitted Gaussian standard deviation, $\sigma$, visualizing the variations.

Taking the results at \SI{0}{\celsius} as an example, the breakdown voltage distribution exhibits good uniformity. In addition, the relationship between the mean of the breakdown voltage ($V_\mathrm{bd,~pad}$) and the temperature obtained from Fig.~\ref{VBDvsT} can be expressed as
\[
V_\mathrm{bd,~pad} = 1.14\ \mathrm{V/K} \times\ T  + 165.9\ \mathrm{V},
\]
indicating that the breakdown voltage increases linearly with temperature. This scaling has been applied to compare measurements performed at different temperatures.

\section{Company Quality Control (CQC)}
\label{CQC}

Breakdown voltage and leakage current properties of each pad of the sensors were obtained on the wafer level.

\subsection{Results of per-pad test}
\label{Per-pad test results}

\begin{figure}[htbp]
	\centering
    \subfigure[]{
	\includegraphics[width=0.47\textwidth]{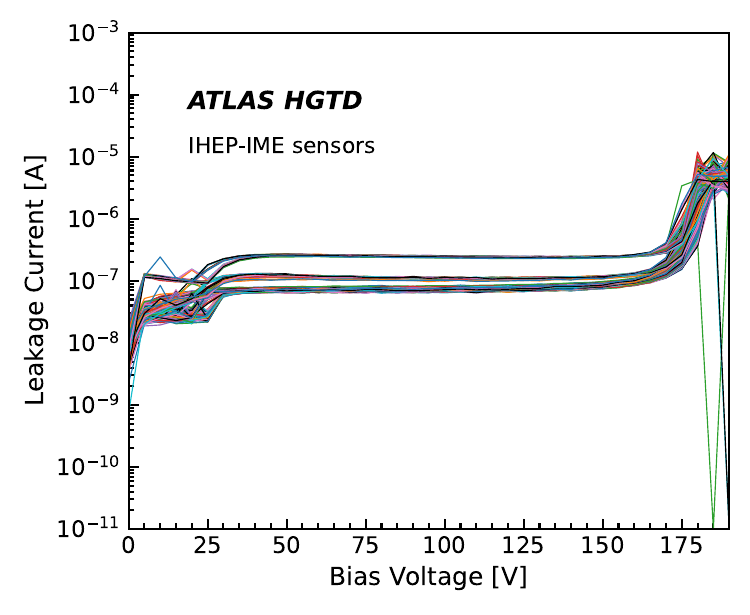}}
    \subfigure[]{
	\includegraphics[width=0.5\textwidth]{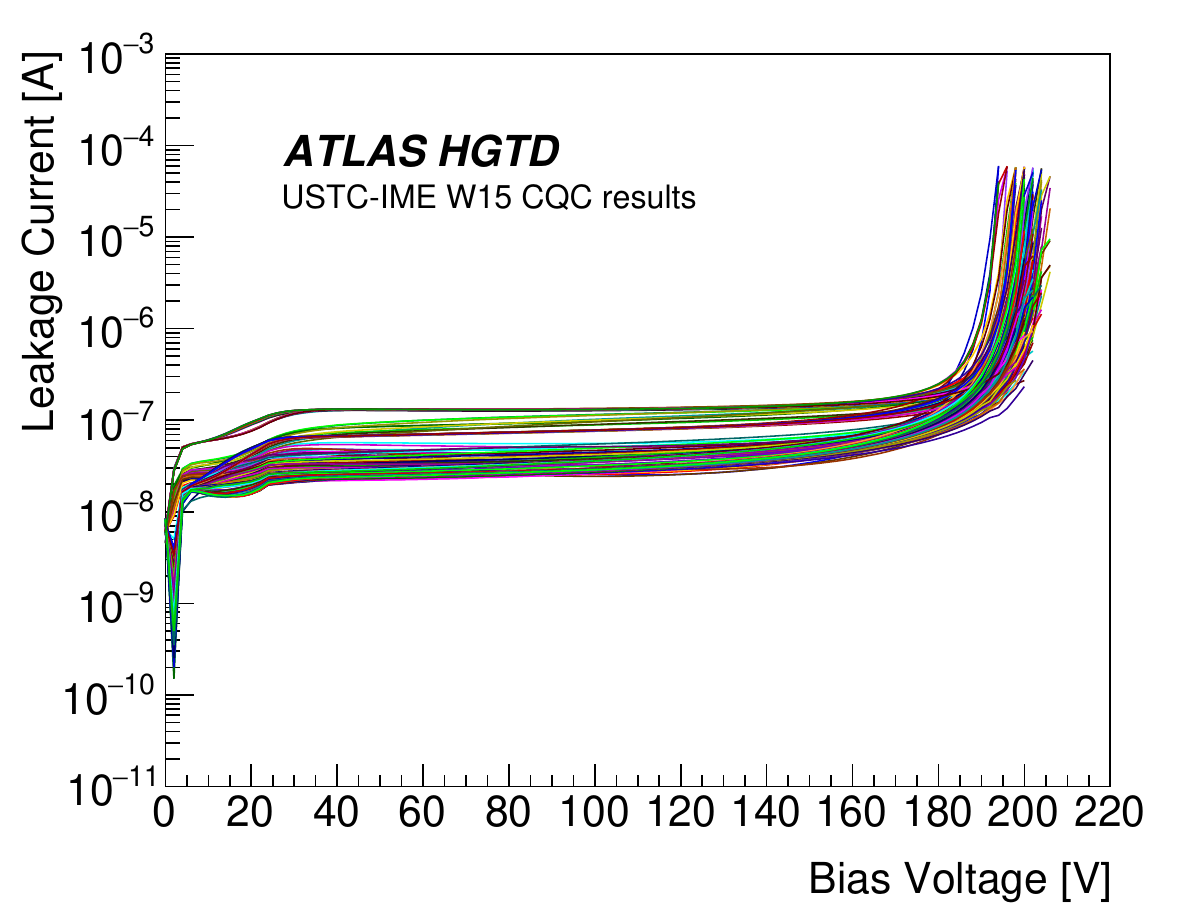}}
    \caption{Example $I-V$ curves obtained from CQC test for an IHEP-IME sensor (a) and a USTC-IME sensor (b).} 
    \label{FIG:CQCIVCURVES}
\end{figure}

$I-V$ measurements were performed on the sensors using probe needles. Example $I-V$ curves are shown for typical sensors in Fig.~\ref{FIG:CQCIVCURVES}.

\begin{figure}[htbp]
	\centering
    \subfigure[]{
	\includegraphics[width=0.45\textwidth]{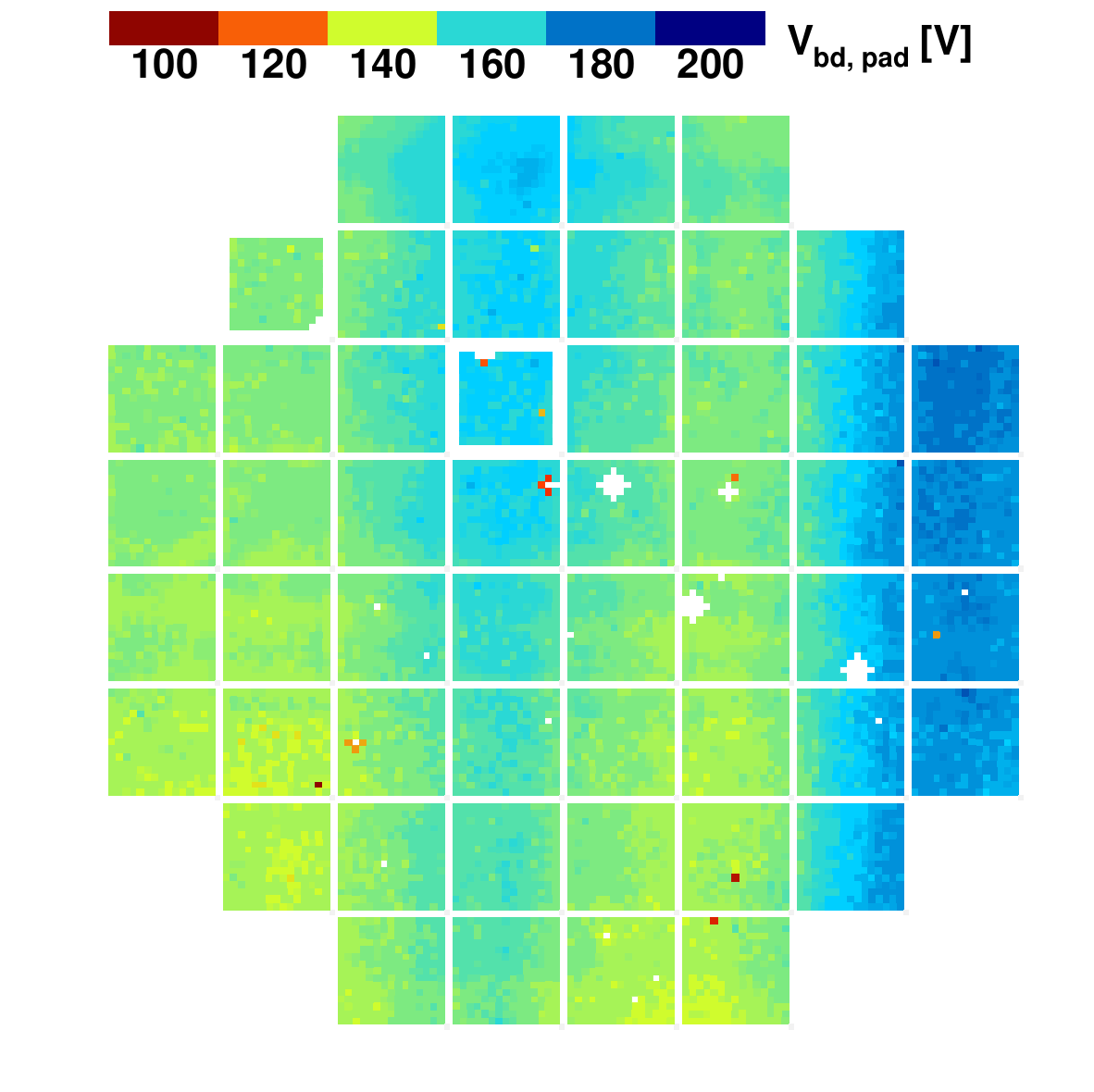}\label{fig:IHEPVbdMap}}
    \subfigure[]{
	\includegraphics[width=0.45\textwidth]{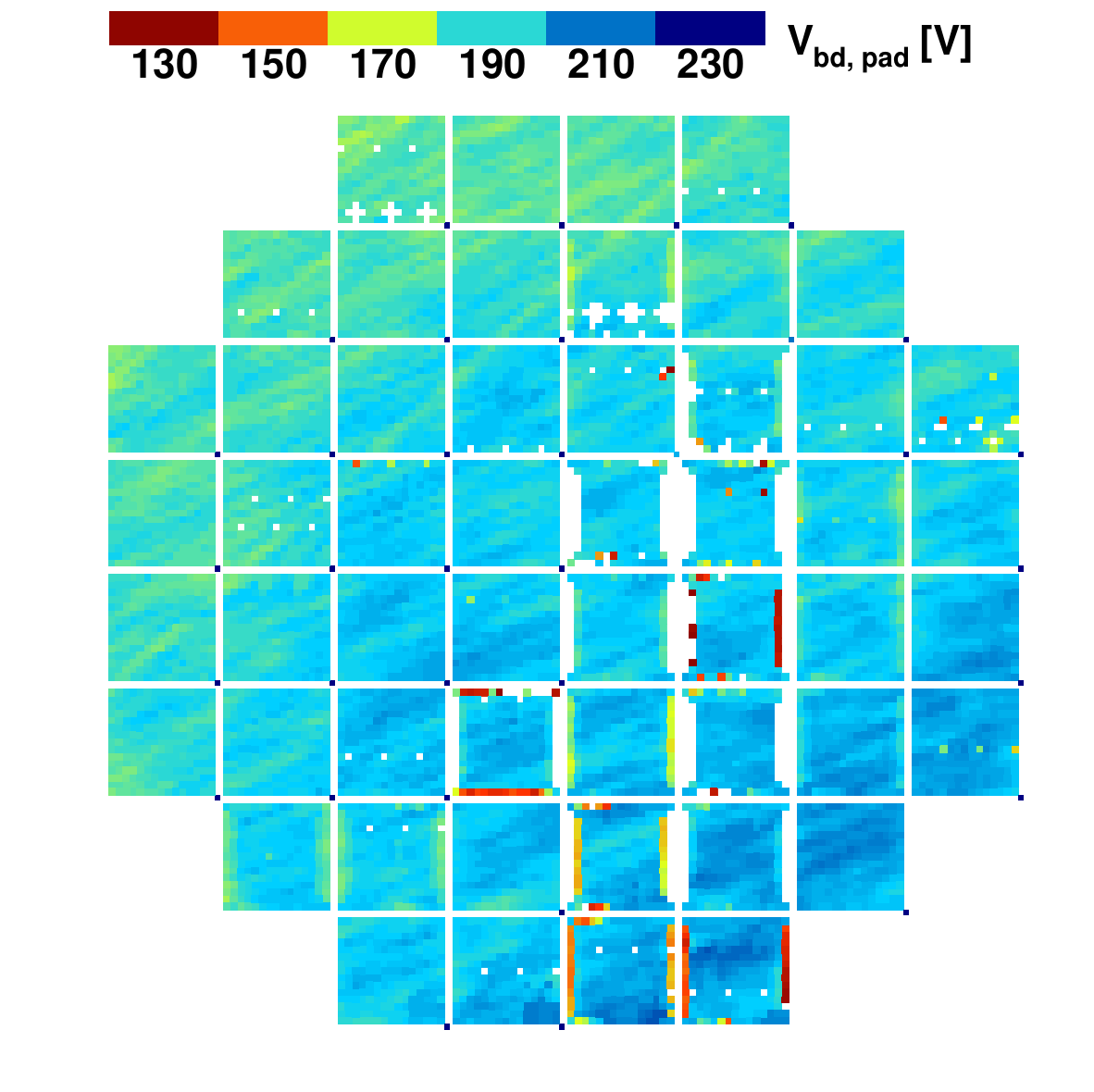}\label{fig:USTCVbdMap}}
    \caption{Distribution of $V\mathrm{_{bd,~pad}}$: (a) A typical IHEP-IME wafer. (b) A typical USTC-IME wafer.}
    \label{vbd distribution on wafer}
\end{figure}

Fig.~\ref{vbd distribution on wafer} shows the $V\mathrm{_{bd,~pad}}$ distribution of two typical wafers, both of which include 52 sensors. There is an extra box in the right bottom corner of each sensor in the USTC-IME figure that represents the GR. The boxes are colored according to the value of $V\mathrm{_{bd,~pad}}$. There is a clear pattern on both maps, believed to originate from the non-uniformity of the gain layer implantation process.

\begin{figure}[htbp]
	\centering
        \subfigure[]{
	    \includegraphics[width=0.45\textwidth]{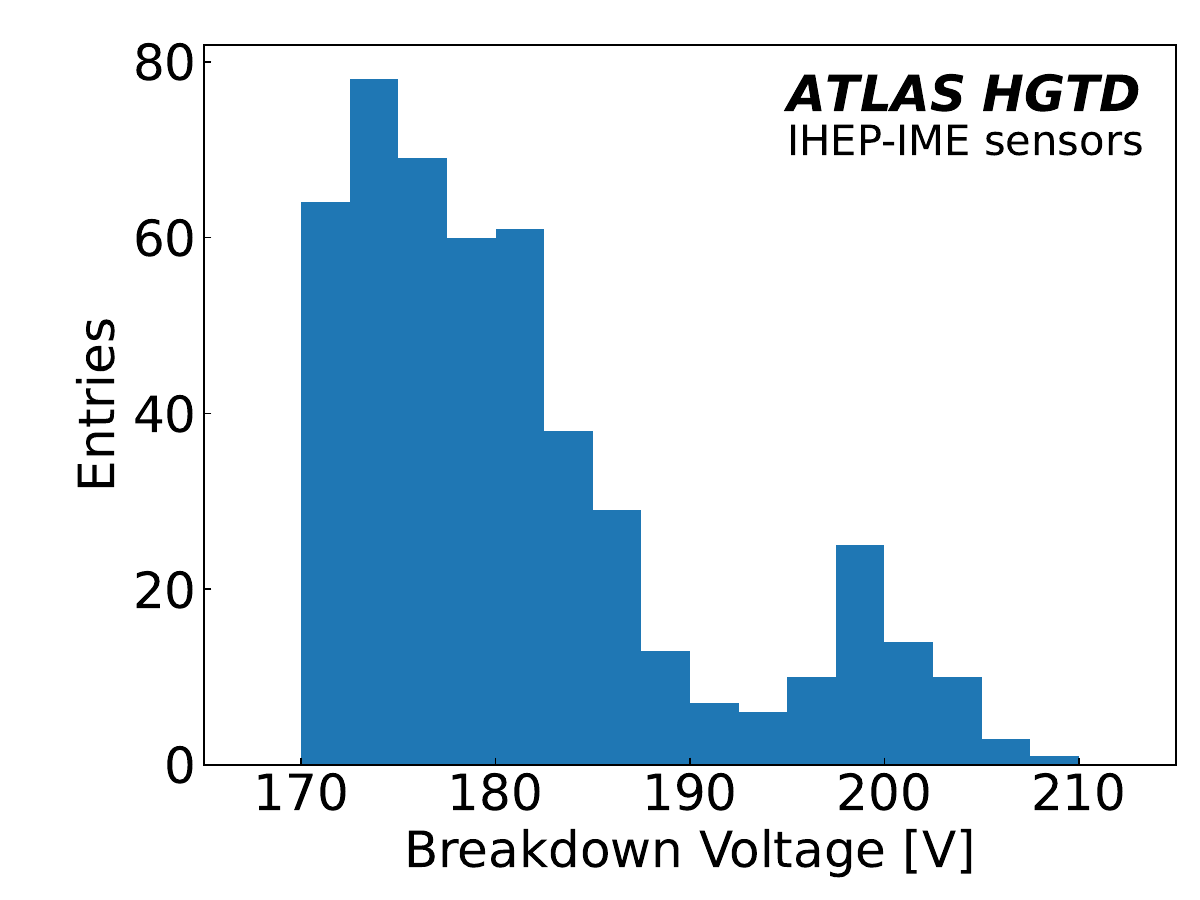}\label{IHPEIMCCQC_passedvbd}}
        \subfigure[]{
	    \includegraphics[width=0.45\textwidth]{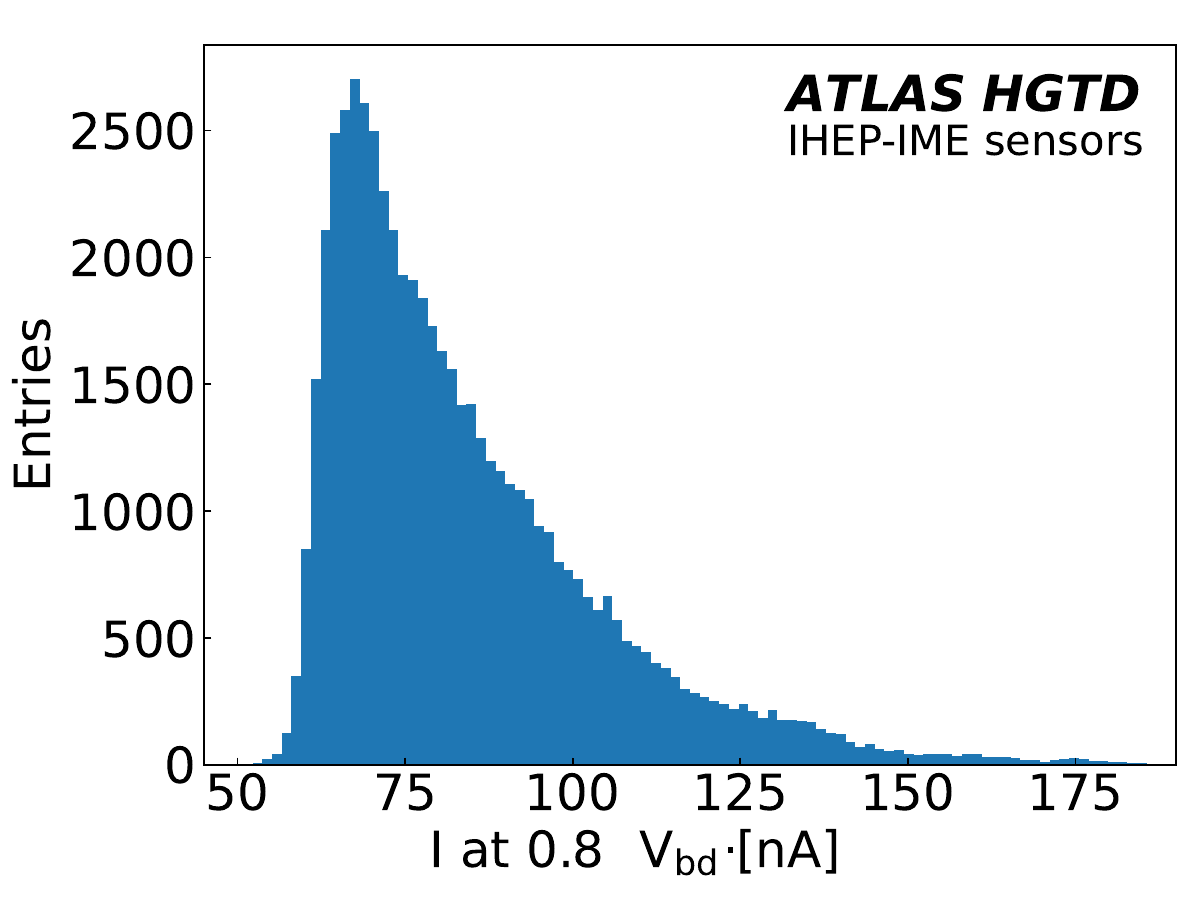}\label{IHEPIMCCQC_passedvbd}}
        \subfigure[]{\includegraphics[width=0.45\textwidth]{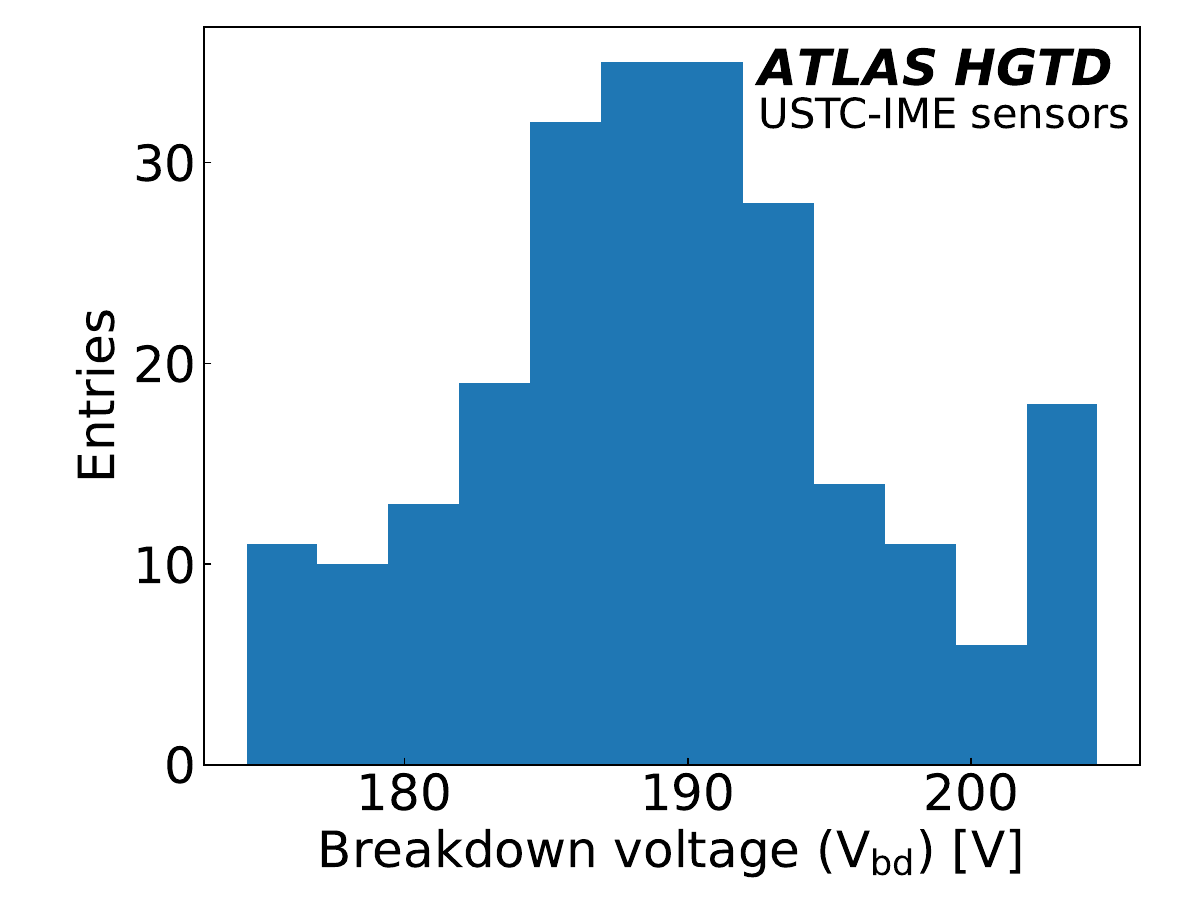}\label{USTCIMCCQC_allpassedvbd}}
        \subfigure[]{\includegraphics[width=0.45\textwidth]{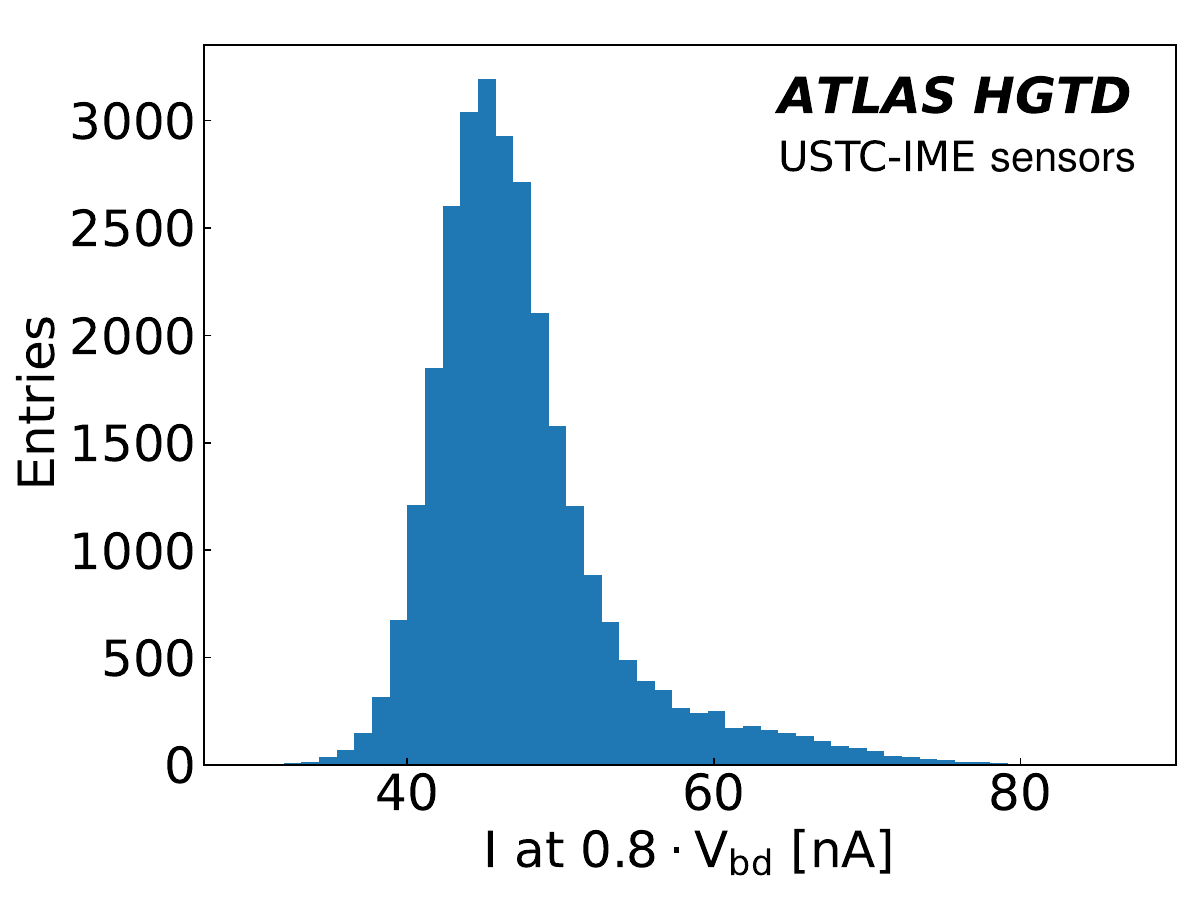}\label{USTCIMCCQC_allpassedi08}}
        \caption{(a) The distribution of breakdown voltage of 488 IHEP-IME sensors.(sensor level) (b) The distribution of per-pad $I$ at 0.8~$\cdot~V_\mathrm{bd}$ of 488 IHEP-IME sensors.  (c) The distribution of $V\mathrm{_{bd}}$ of 232 USTC-IME sensors.  (d) The distribution of per-pad $I$ at 0.8~$\cdot~V_\mathrm{bd}$ of 232 USTC-IME sensors. The temperature correction (1.14 V/K) was not applied here.}
        \label{CQCDataAnalysis}
\end{figure}

\begin{figure}[htbp]
	\centering
	\includegraphics[width=0.45\textwidth]{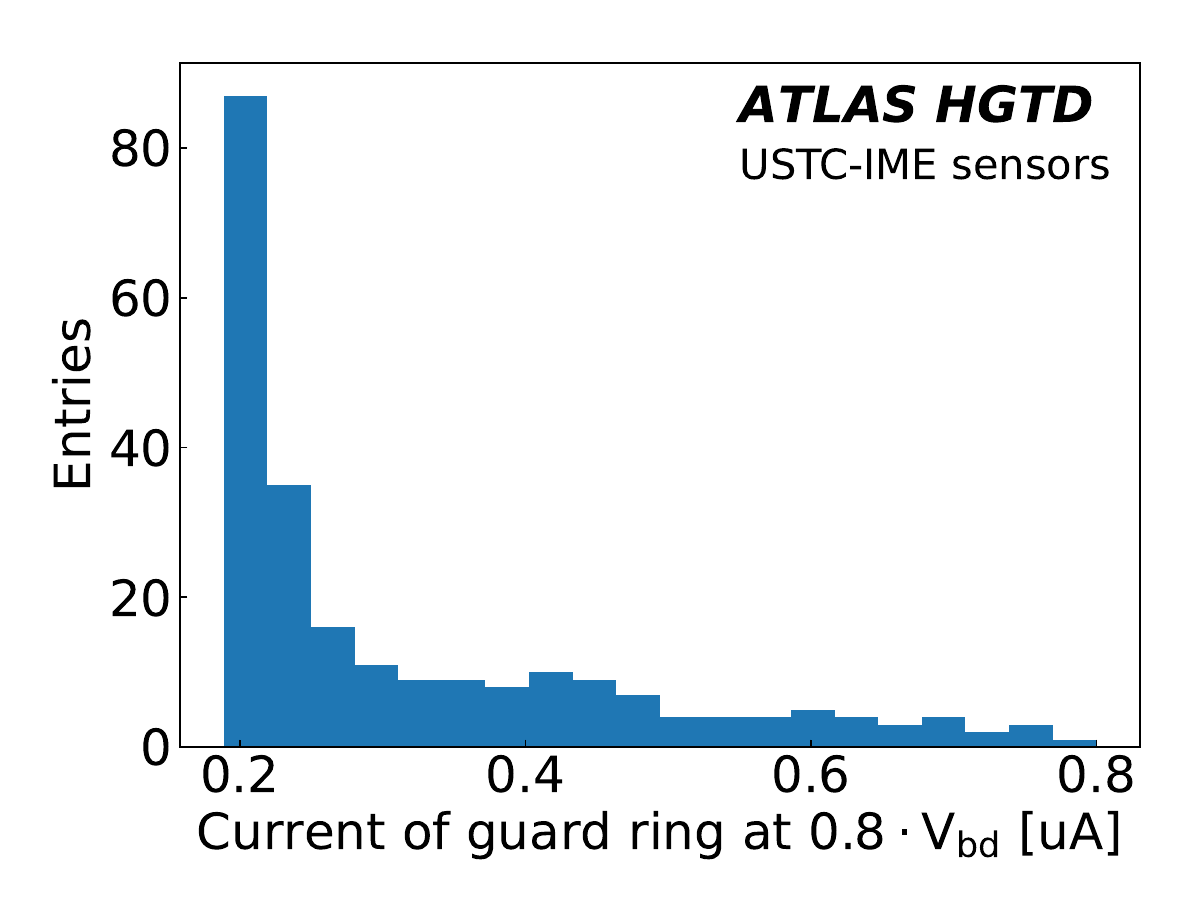}
    \caption{The distribution of $I$ at 0.8$\cdot V_\mathrm{bd}$ of guard ring of USTC-IME sensors.}
    \label{USTCCQCDataAnalysisGR}
\end{figure}

In the pre-production batch, 488 IHEP-IME sensors and 232 USTC-IME sensors meet the requirements. The distributions of the breakdown voltage and leakage current of these sensors are shown in Fig.\ref{CQCDataAnalysis}. The distributions of the sensors' $V\mathrm{_{bd}}$ are shown in Fig.~\ref{IHPEIMCCQC_passedvbd} and~\ref{USTCIMCCQC_allpassedvbd}. The $I$ at 0.8~$\cdot~V_\mathrm{bd}$ of the pads, excluding the pads in the outer two rings, is 30 to 190 nA (see Fig.~\ref{IHEPIMCCQC_passedvbd} and~\ref{USTCIMCCQC_allpassedi08}). In addition, the guard rings' current at 0.8~$\cdot~V\mathrm{_{bd}}$ of USTC-IME good sensors is in the range of 100 to 800 nA, shown in Fig.~\ref{USTCCQCDataAnalysisGR}.

\begin{figure}[htbp]
	\centering
    \subfigure[]{
	\includegraphics[width=0.45\textwidth]{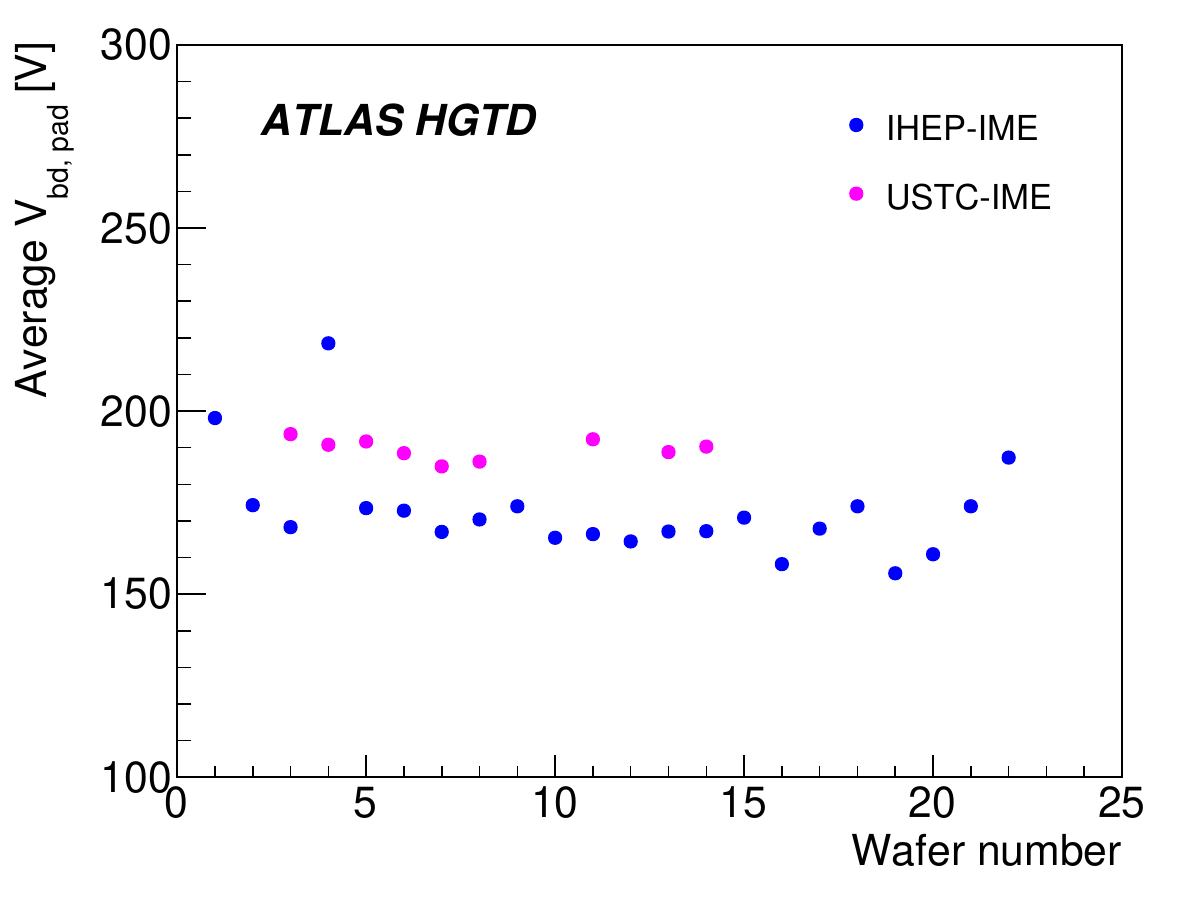}\label{fig:Vbd_merged}}
    \subfigure[]{
	\includegraphics[width=0.45\textwidth]{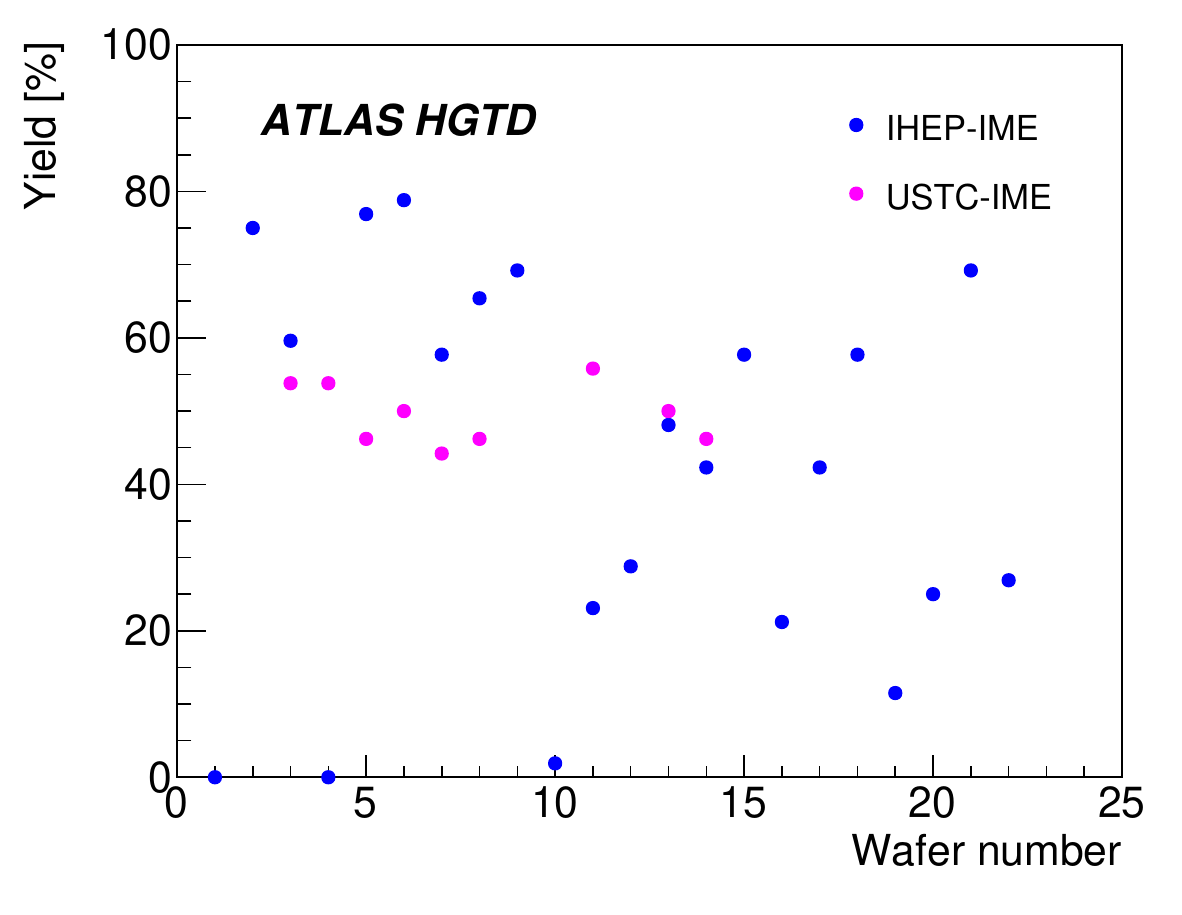}\label{fig:Yield_merged}}
    \caption{ (a) The average $V_\mathrm{bd,~pad}$ of IHEP-IME and USTC-IME wafers. (b) The yield of IHEP-IME and USTC-IME wafers.}
    \label{fig:CQCsummary}
\end{figure}

The absolute yields can not be quantified in this study since data are available only for wafers with at least 18 good sensors for the USTC-IME production, and for wafers in a pre-selected $\mathrm{V_{bd}}$ range for the IHEP-IME production. Within the available data, the average $V_\mathrm{bd,~pad}$ and yield of the IHEP-IME wafers and USTC-IME wafers are summarized in Fig.~\ref{fig:CQCsummary}. The sensor yield of IHEP-IME pre-production varies for different wafers and the sensor yield of USTC-IME pre-production fluctuates around 50$\%$. There are two IHEP-IME wafers with 0$\%$ yield, due mostly to the high $\mathrm{V_{bd}}$. 

\subsection{Results of 15 \textit{×} 1 probe-card test}
\label{CQC15x1results}

\begin{figure}[htbp]
	\centering
    \subfigure[]{
	\includegraphics[width=0.45\textwidth]{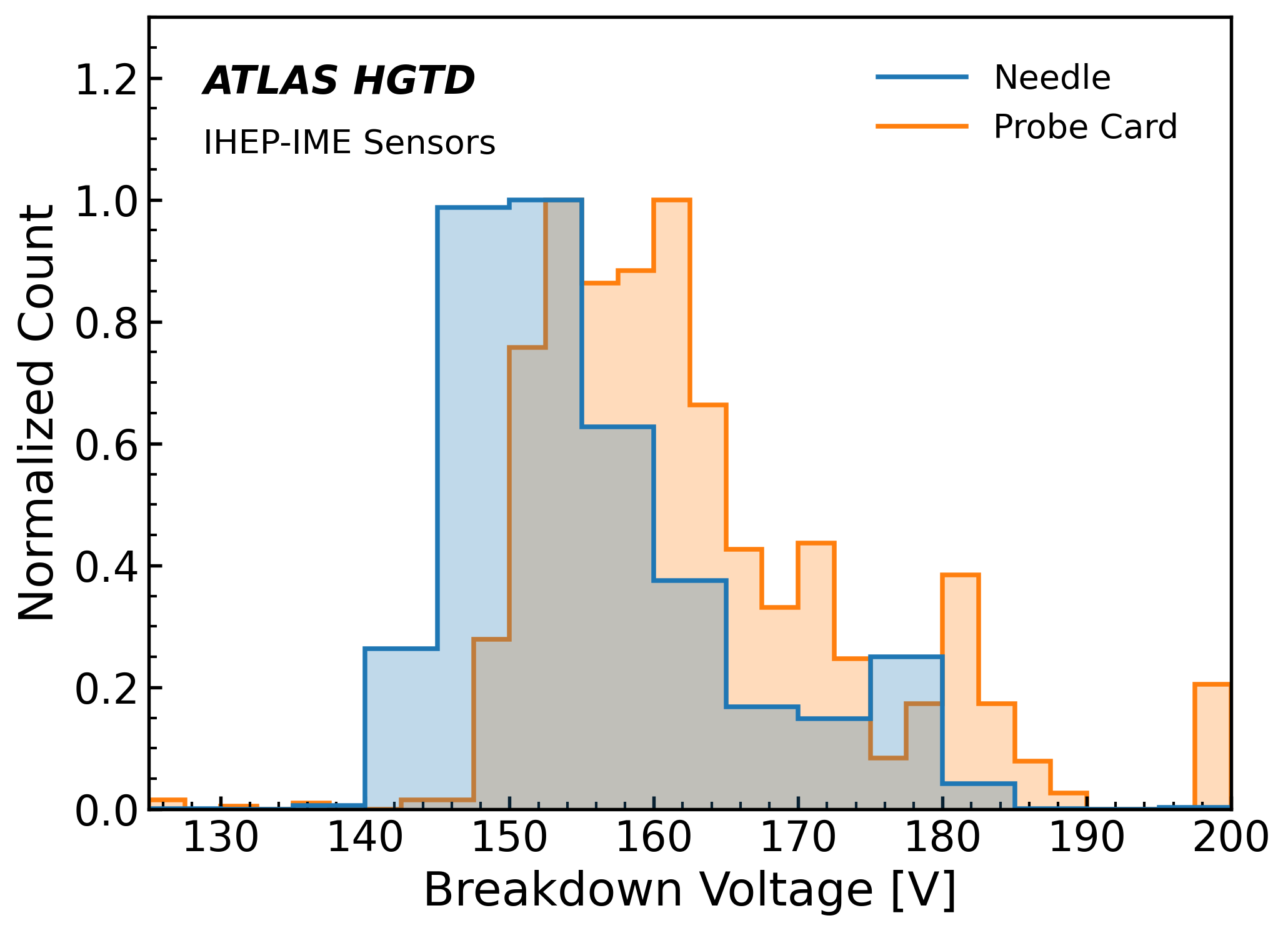}\label{fig:cqcvbdcom}}
    \subfigure[]{
	\includegraphics[width=0.45\textwidth]{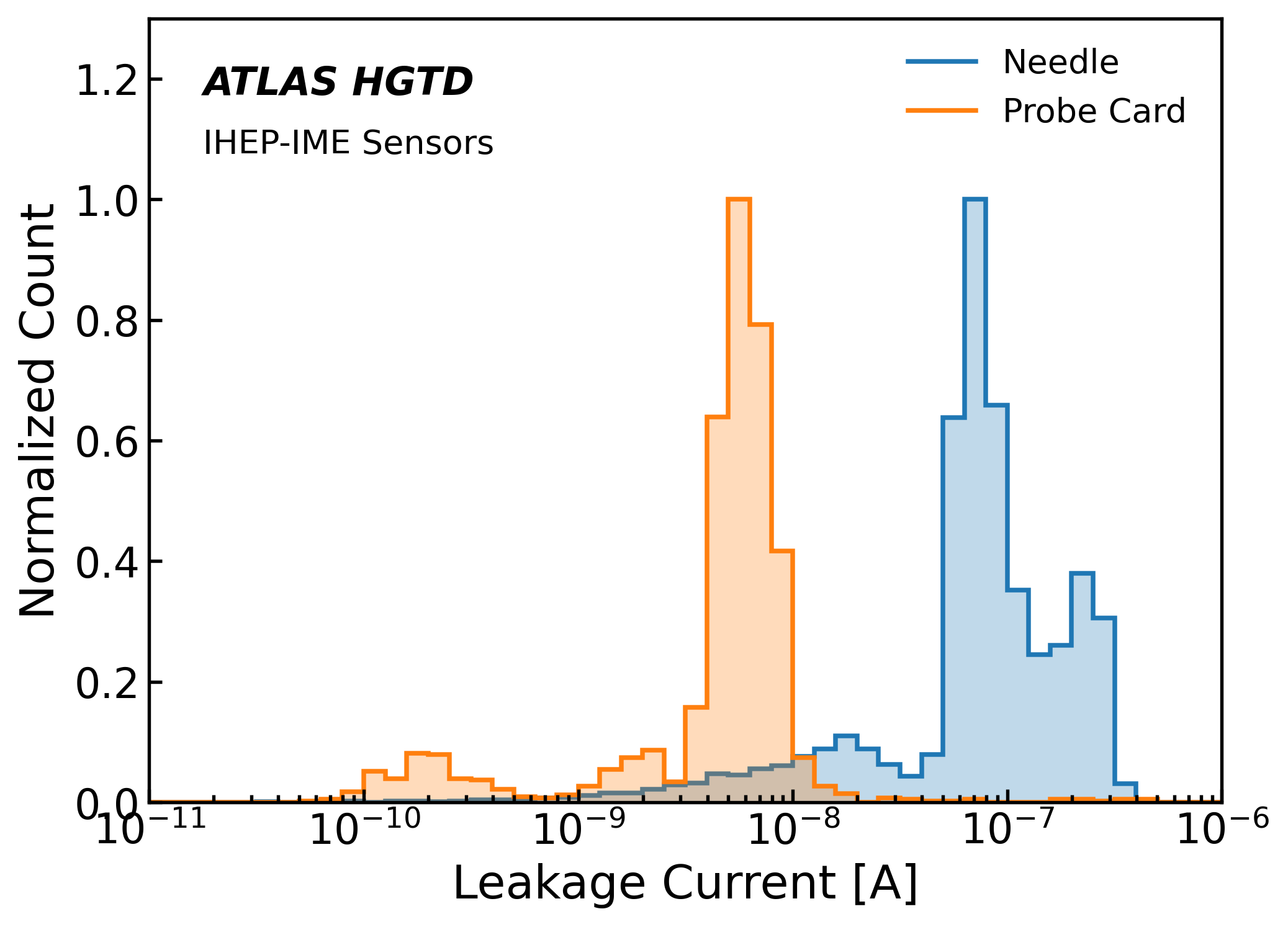}\label{fig:cqccurrentcom}}
    \caption{(a) Distribution of breakdown voltages of IHEP-IME sensors measured by the probe needle and probe card methods. (b) Distribution of leakage current of IHEP-IME sensors.  A total of 104 sensors are tested, and each sensor contains 15 ~$\times$~15 pads. The probe needle test is performed pad by pad, corresponding to 104~$\times$~225 pad-level measurements, while the probe card test was performed column by column, corresponding to 104~$\times$~15 column-level measurements. The histograms are normalized to the largest bin for shape comparison. The leakage current measured by the probe card was divided by 15 for comparison with pad-level values.}
    \label{fig:pcsensorIV}
\end{figure}

To evaluate the reliability of the $15~\times~1$ probe-card measurement method, two complete IHEP-IME wafers, corresponding
to 104 sensors, were characterized using both the single-needle and the $15~\times~1$ probe-card setups. For a direct comparison of the breakdown-voltage observable, the single-needle $I$-$V$ data were first combined at the column level. At each bias voltage, the currents of the 15 pads belonging to the same column were summed. The breakdown voltage extracted from this column-summed current is denoted by $V_{\mathrm{bd,15pad}}$ and is defined as the voltage at which the summed current reaches \SI{7.5}{\micro A}, corresponding to 15 times the single-pad threshold of \SI{500}{\nano A}. The same definition is applied to the probe-card data, for which the 15 pads in one column are contacted and measured simultaneously. Therefore, the two distributions compared in Fig.~\ref{fig:cqcvbdcom} represent the same column-level observable, $V_{\mathrm{bd,15pad}}$, obtained using the two measurement methods.

\begin{figure}[htbp]
	\centering
    \subfigure[]{
	\includegraphics[width=0.45\textwidth]{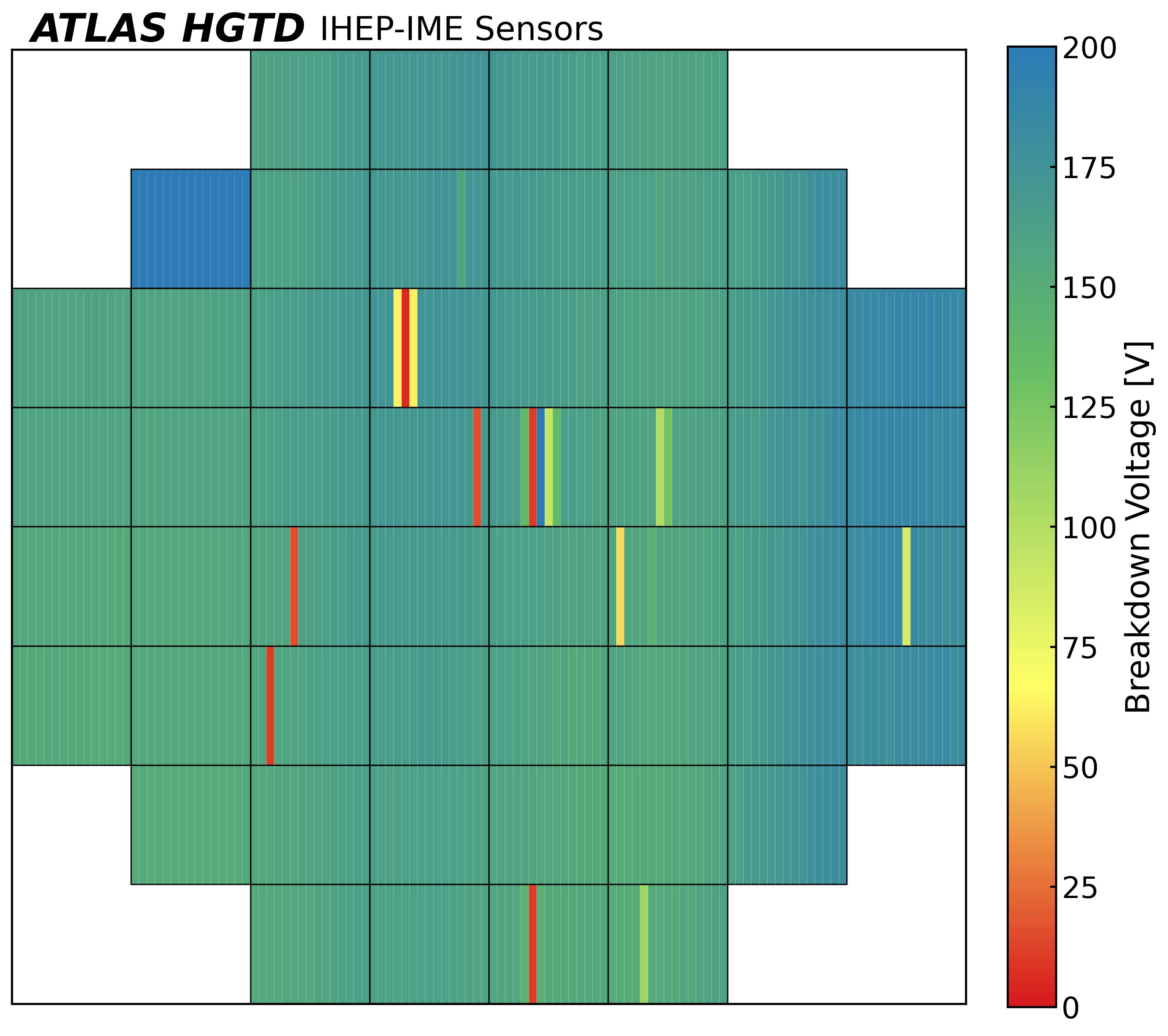}\label{fig:CQCPCVbdMap}}
    \subfigure[]{
	\includegraphics[width=0.45\textwidth]{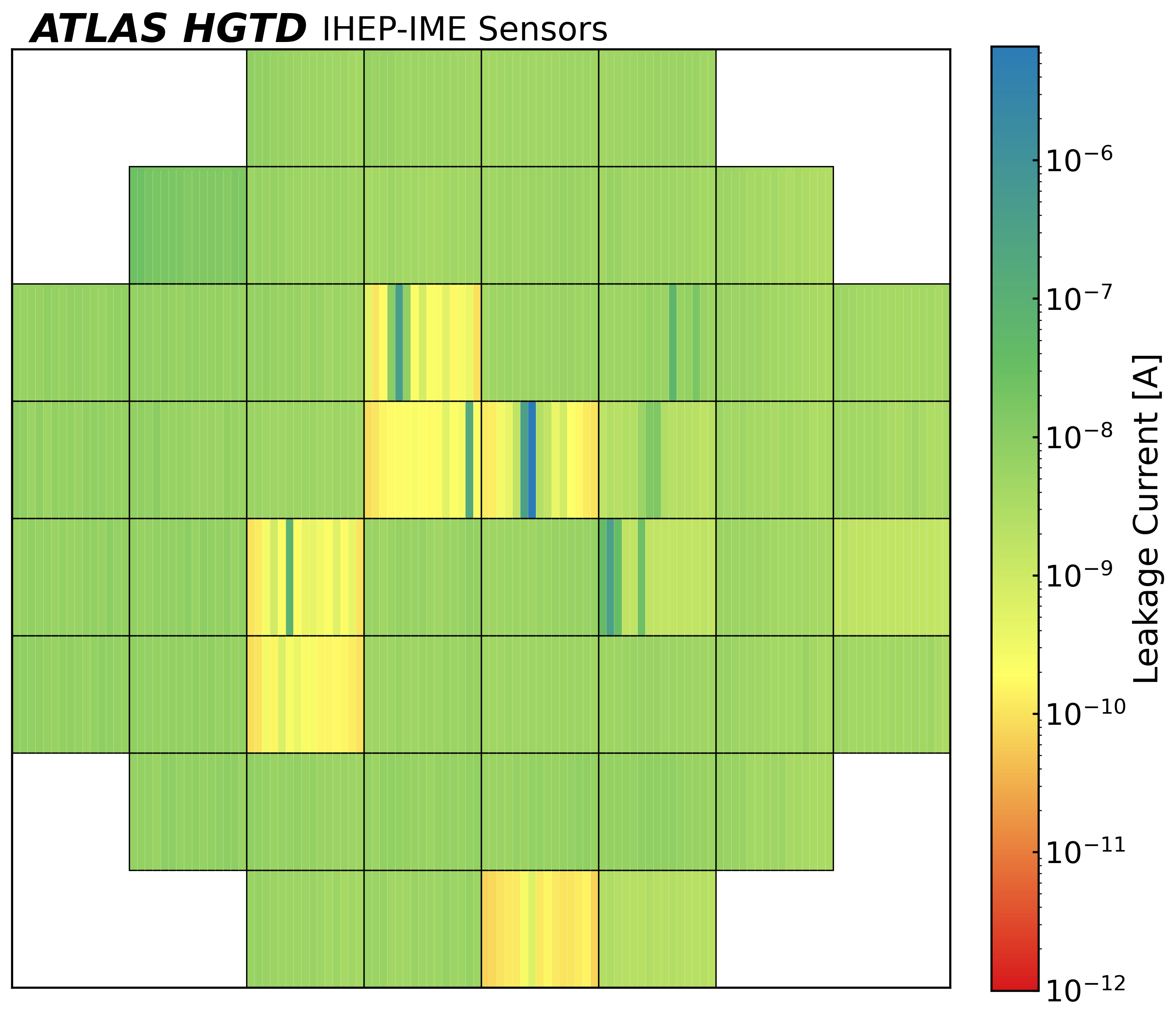}\label{fig:CQCPCKGMap}}
    \caption{ (a) Breakdown voltage distribution of the IHEP-IME wafer 21, measured by 15~$\times$~1 probe card. (b) Leakage current distribution of the same wafer, measured by 15~$\times$~1 probe card. }
    \label{fig:pcwaferIV}
\end{figure}

The 15~$\times$~1 probe-card maps for IHEP-IME wafer 21 are shown in Fig.~\ref{fig:pcwaferIV}. Figure~\ref{fig:cqcvbdcom} compares the normalized $V_{\mathrm{bd,15pad}}$ distributions obtained from the column-summed single-needle data and from the directly measured probe-card data. Each distribution contains $104\times15$ column-level entries. For the leakage-current comparison in Fig.~\ref{fig:cqccurrentcom}, the individual-pad currents from the single-needle measurement are compared with the probe-card column currents divided by 15, providing a per-pad-equivalent current scale. The needle and probe-card leakage-current distributions therefore contain $104\times225$ pad-level entries and $104\times15$ column-level entries, respectively. Each histogram is normalized to its own largest bin for comparison of the
distribution shapes. All 225 pads per sensor are included in these diagnostic distributions; the exclusion of the outermost two pad rings applies only to the leakage-current uniformity
acceptance observable defined in Section~\ref{Acceptance criteria}.

A systematic difference in the measured leakage-current scale is observed between the two methods. This difference is mainly attributed to the different guard-ring connection schemes. In the single-needle per-pad measurement, the guard ring is left floating, whereas in the $15~\times~1$ probe-card measurement it is grounded. Differences in the contact configuration and acquisition settings may also contribute to the observed offset. To retain a conservative sensor selection in the presence of this setup-dependent difference and the summation over 15 pads, the leakage-current uniformity criterion is tightened for the
IHEP-IME CQC performed with the $15~\times~1$ probe card, as described in Section~\ref{Acceptance criteria}. The validated
$15~\times~1$ probe-card method is therefore adopted for the IHEP-IME production CQC, reducing the number of probe positions per sensor from 225 to 15 and substantially reducing the measurement time.

\subsection{Spatial yield analysis}

\begin{figure}[htbp]
	\centering
    \subfigure[]{
	\includegraphics[width=0.45\textwidth]{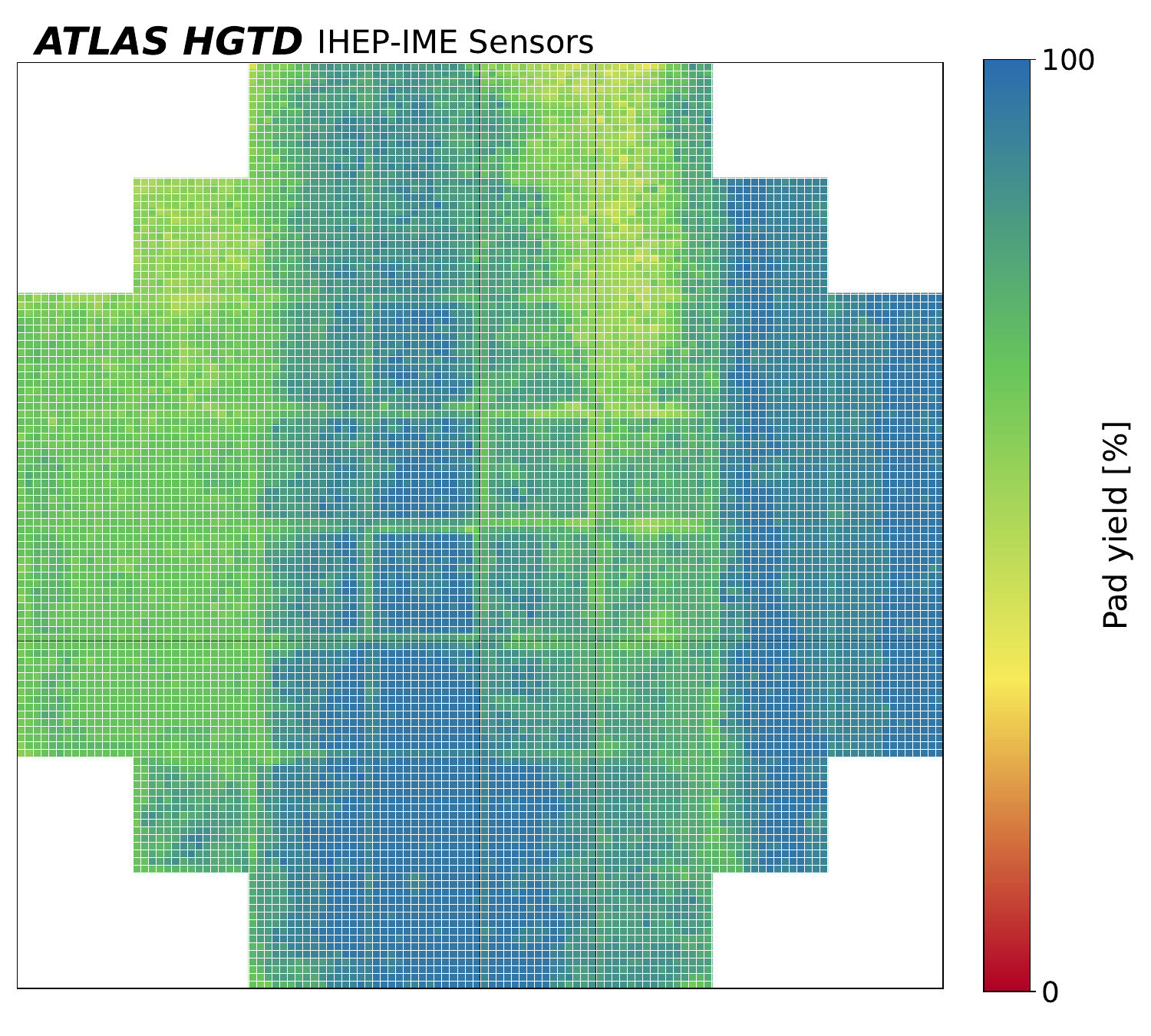}\label{fig:yieldofpad}}
    \subfigure[]{
	\includegraphics[width=0.5\textwidth]{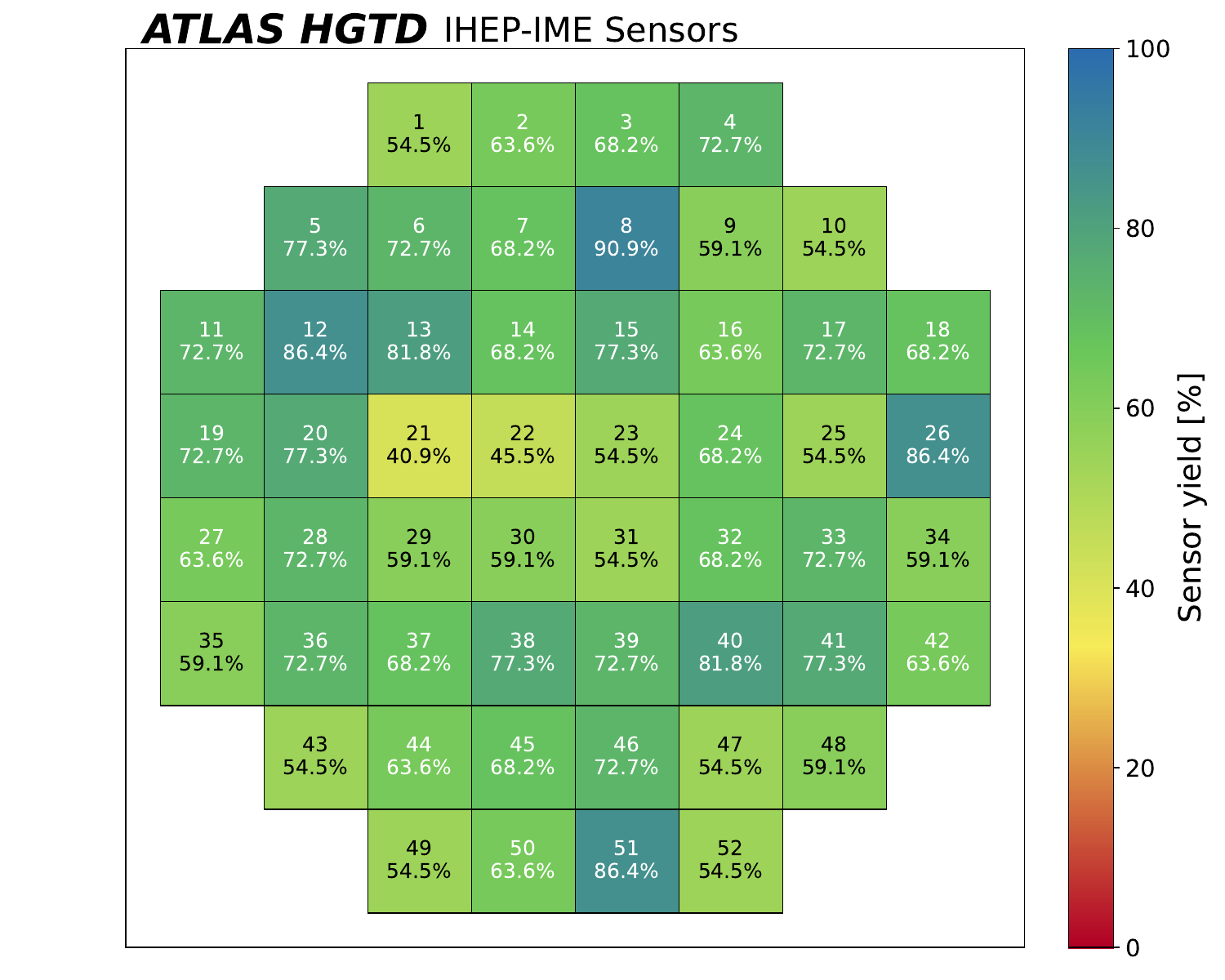}\label{fig:yieldofsensor}}
    \caption{ (a) The pad yield of 22 IHEP-IME wafers. (b) The sensor yield of 22 IHEP-IME wafers. Darker shades indicate higher yield, while lighter shades indicate lower yield. }
    \label{fig:Yieldwafer}
\end{figure}

As shown in section~\ref{Per-pad test results}, the distribution of $V\mathrm{_{bd,~pad}}$ on the wafer exhibits a dependence on position. Therefore, the spatial distribution of the yield on the IHEP-IME wafers is analyzed. Fig.~\ref{fig:yieldofpad} shows the yield of pads, where one box represents one pad. In each block of Fig.~\ref{fig:yieldofsensor}, the integer number represents the sensor position on the wafer, and the yield is listed below. The distribution of the sensor yield does not show a significant position dependence, but the spatial distribution of the pad yield shows an obvious correlation with the spatial distribution of the sensor yield, as expected. 

\section{Sensor Quality Control (SQC)}
\label{SQC}
SQC includes the per-pad tests under the same conditions as CQC for selected IHEP-IME sensors and the 15~$\times$~15 matrix tests for selected USTC-IME sensors. The per-pad test is essential to verify the results of the CQC and also to verify that UBM and dicing don’t have significant effect. The 15~$\times$~15 matrix test can perform the electrical characterization of the sensor with all pads grounded, which is consistent with the actual operating state of the sensor (except for the floating of the GR during the SQC). The per-pad measurement is not done for USTC-IME sensors during the SQC because it has been done for all pre-production sensors during the CQC at the wafer level. 

\subsection{Results of per-pad tests}

\begin{figure}[htbp]
	\centering
    \subfigure[]{
	\includegraphics[width=0.45\textwidth]{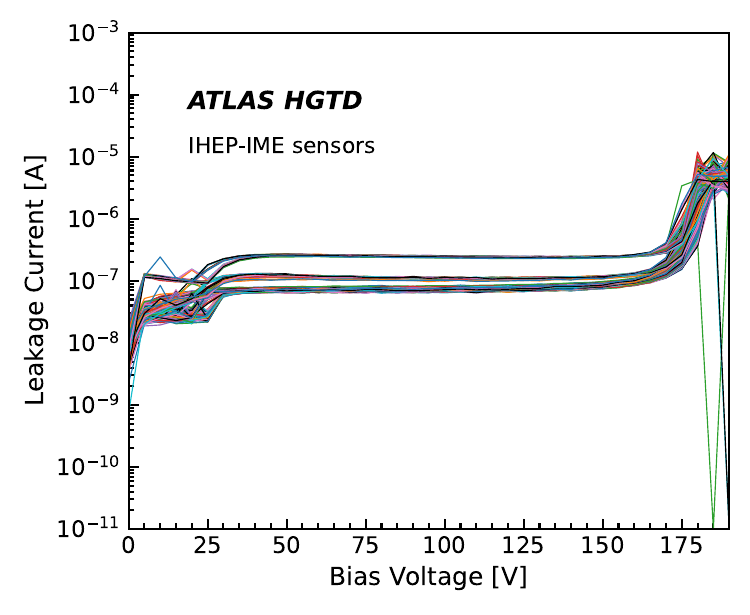}}
    \subfigure[]{
	\includegraphics[width=0.45\textwidth]{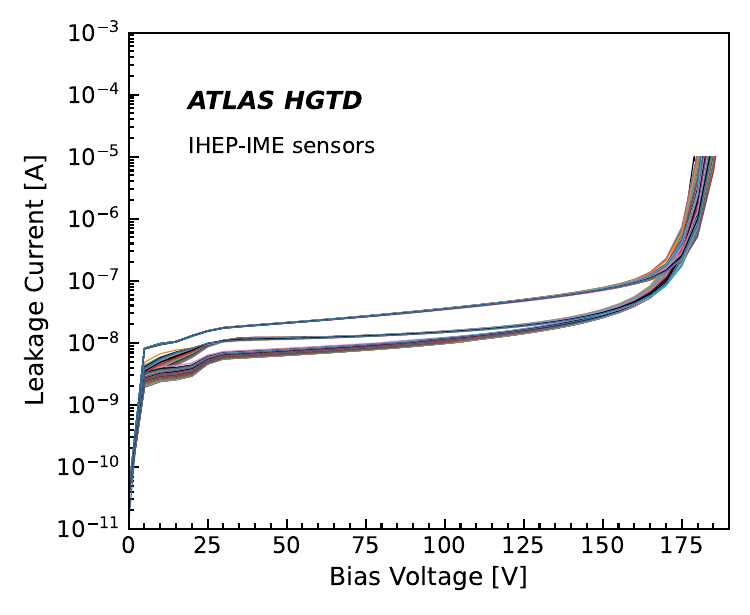}}
    \subfigure[]{
	\includegraphics[width=0.45\textwidth]{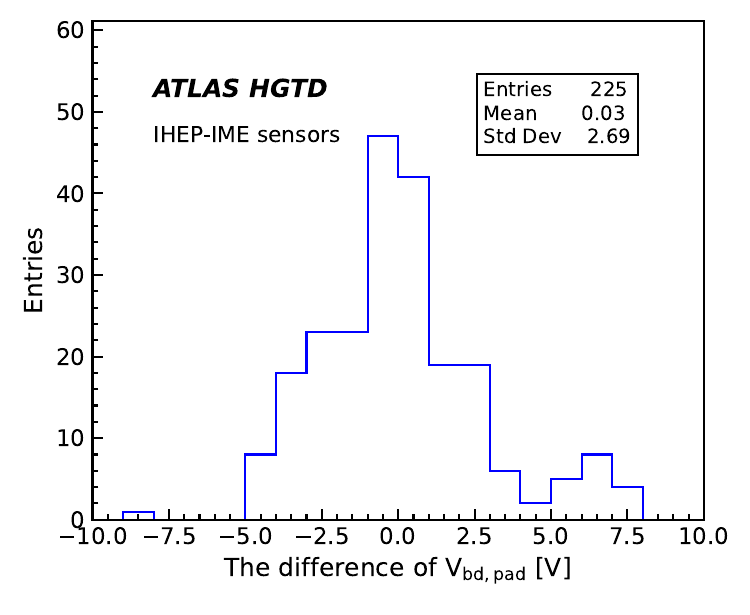}}
    \subfigure[]{
	\includegraphics[width=0.45\textwidth]{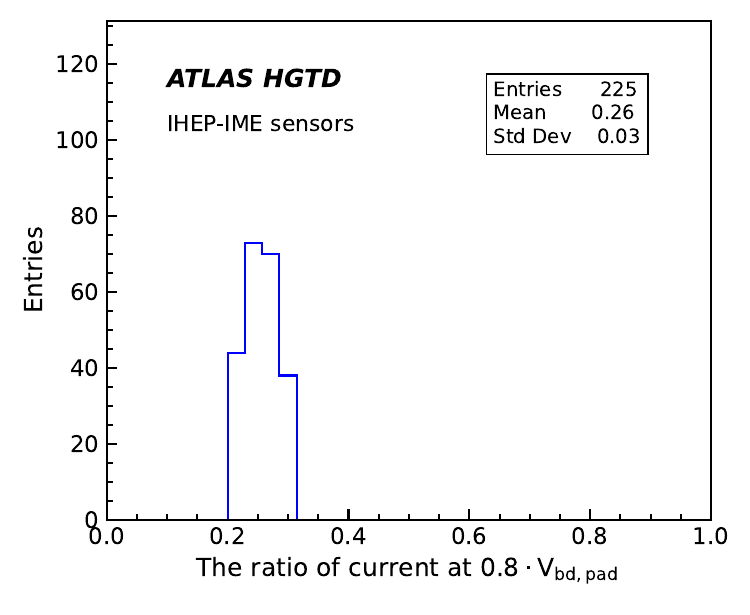}}
    \caption{Per-pad CQC--SQC comparison for an IHEP-IME sensor.
(a) Wafer-level CQC $I-V$ curves before UBM deposition and dicing.
(b) SQC $I-V$ curves after UBM deposition and dicing.
(c) Distribution of
$\Delta V_{\mathrm{bd,pad}}
=V_{\mathrm{bd,pad}}^{\mathrm{CQC}}
-V_{\mathrm{bd,pad}}^{\mathrm{SQC}}$.
(d) Distribution of
$r_I=
I^{\mathrm{CQC}}(0.8\,V_{\mathrm{bd,pad}})
/
I^{\mathrm{SQC}}(0.8\,V_{\mathrm{bd,pad}})$.}
    \label{IHEPPerpadlevelcomparison}
\end{figure}

Figure~\ref{IHEPPerpadlevelcomparison} compares one IHEP-IME
sensor measured at wafer level during CQC and remeasured after UBM
deposition and dicing during SQC. The breakdown-voltage difference,

\begin{equation}
\Delta V_{\mathrm{bd,pad}}
=
V_{\mathrm{bd,pad}}^{\mathrm{CQC}}
-
V_{\mathrm{bd,pad}}^{\mathrm{SQC}},    
\end{equation}

has a mean of \SI{0.03}{\volt} and a standard deviation of \SI{2.69}{\volt}. The mean is consistent with zero, and no systematic displacement of the breakdown-voltage distribution is observed. This supports the CQC determination of $V_{\mathrm{bd,pad}}$. In this example, no new localized early-breakdown feature is observed after UBM deposition and dicing.

The leakage-current ratio,

\begin{equation}
r_I
=
\frac{
I^{\mathrm{CQC}}(0.8\,V_{\mathrm{bd,pad}})
}{
I^{\mathrm{SQC}}(0.8\,V_{\mathrm{bd,pad}})
},
\end{equation}

has a mean of 0.26 and a standard deviation of 0.03. Thus, the absolute SQC currents are systematically higher than the CQC currents, although the distribution is narrow across the 225 pads. The measurements were performed at different processing stages and with different acquisition settings, and the available data do
not isolate a unique cause of this scale offset. The narrow ratio distribution indicates a systematic offset rather than a small number of newly damaged pads. The CQC--SQC validation is therefore based on the breakdown-voltage consistency and the absence of new local outliers, rather than the consistency of absolute leakage currents.

\subsection{Results of $I-V$ and $C-V$ tests with the 15 \textit{×} 15 probe card}

\begin{figure}[htbp]
	\centering
    \subfigure[]{\includegraphics[width=0.45\textwidth]{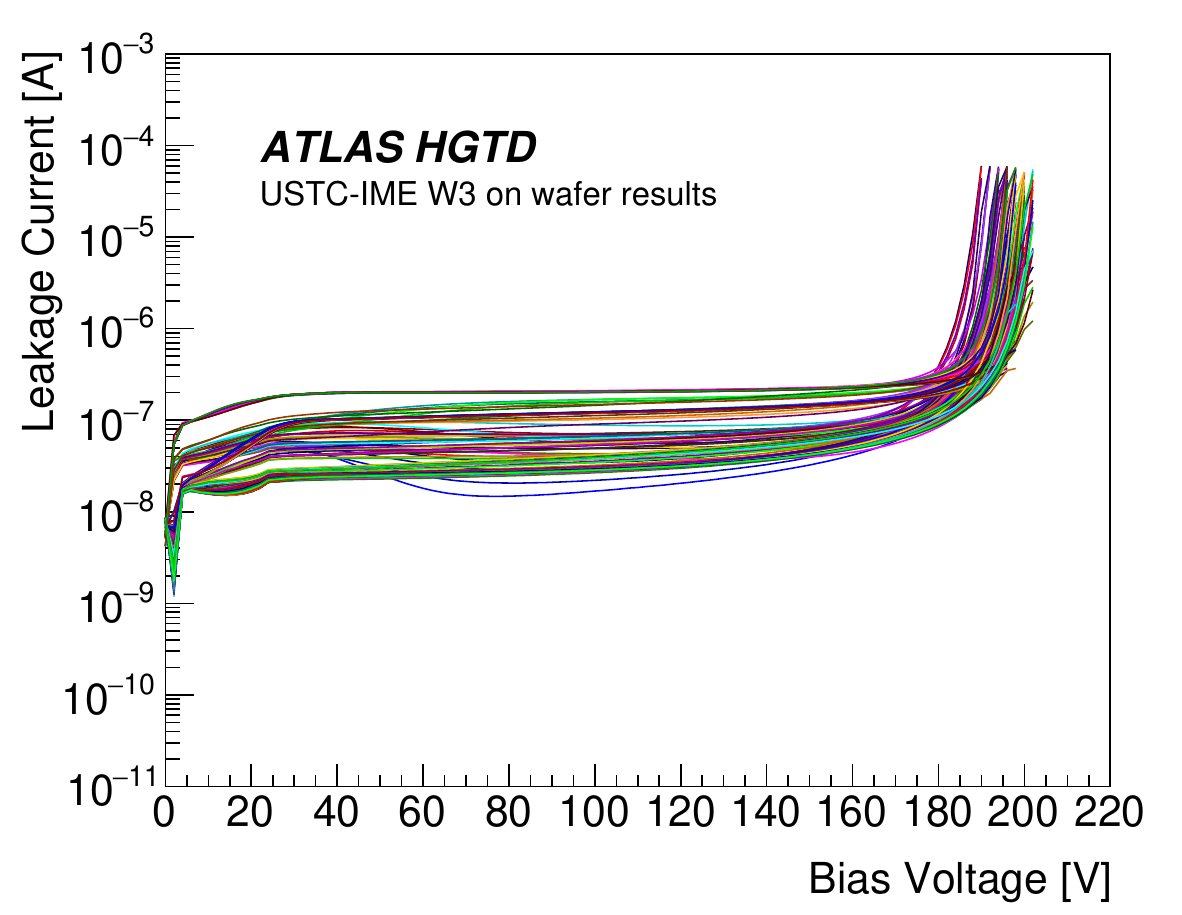}\label{USTCW3P5IVneedles}}
    \subfigure[]{\includegraphics[width=0.45\textwidth]{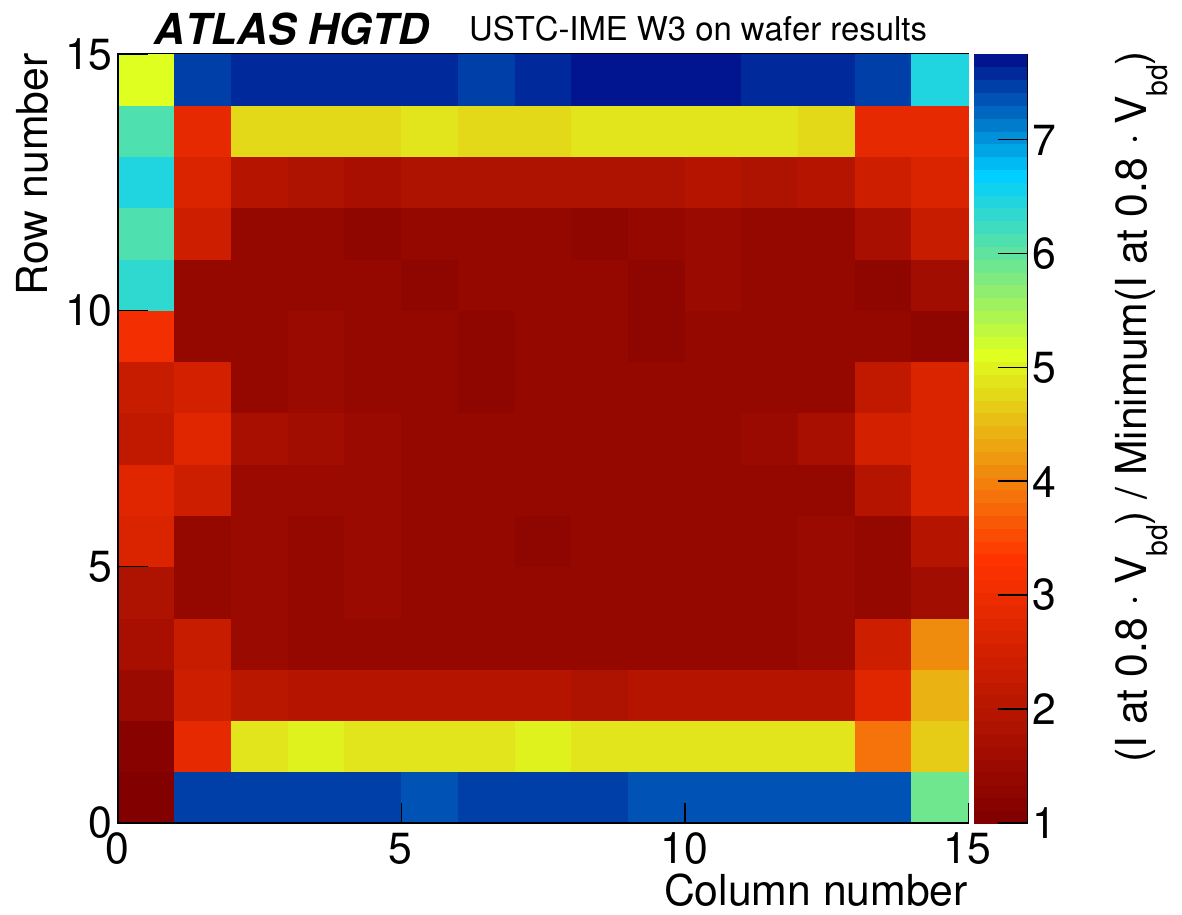}\label{USTCW3P5OnWaferCR_UBM}}
    \subfigure[]{
	\includegraphics[width=0.45\textwidth]{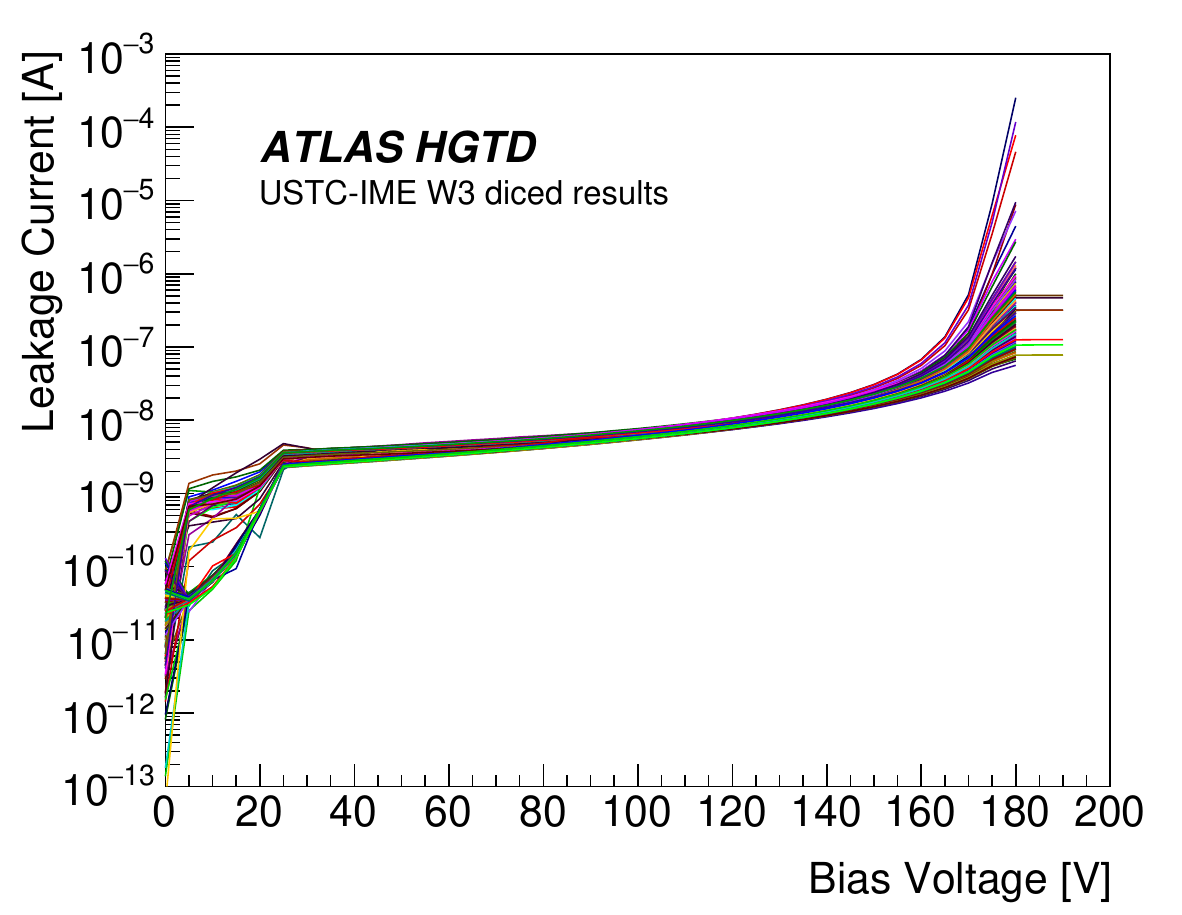}\label{USTCW3P5IV15x15matrix}}
    \subfigure[]{
	\includegraphics[width=0.45\textwidth]{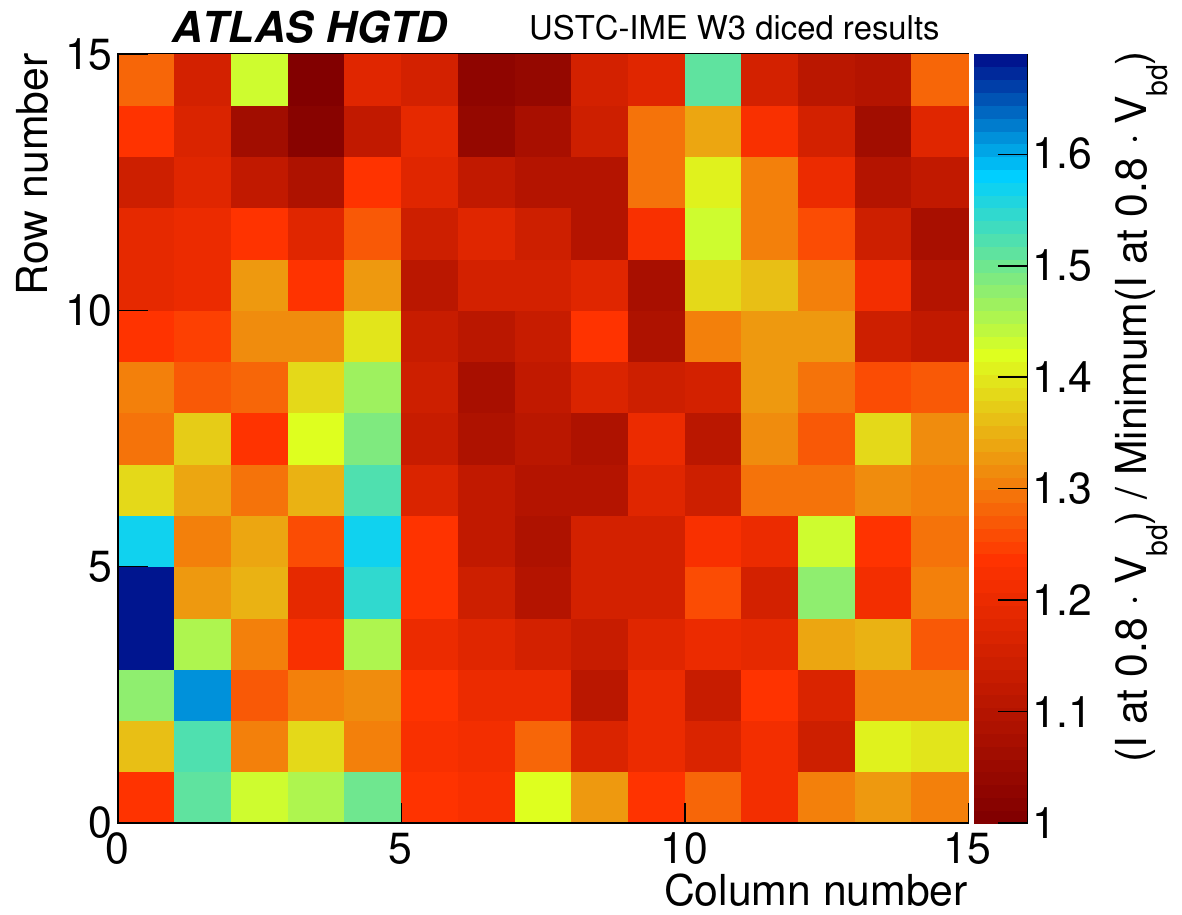}\label{USTCW3P5DicedCR_UBM}}
    \caption{The $I-V$ results measured by probe needles and 15~$\times$~15 probe card of a USTC-IME sensor. (a) The $I-V$ curves tested by the probe needles at room temperature. (b) The ($I$ at 0.8~$\cdot~V_\mathrm{bd}$)/Min($I$ at 0.8~$\cdot~V_\mathrm{bd}$) tested by the probe needles at room temperature. (c) The $I-V$ curves tested by the 15~$\times$~15 probe card at \SI{20}{\celsius}. (d) The ($I$ at 0.8~$\cdot~V_\mathrm{bd}$)/Min($I$ at 0.8~$\cdot~V_\mathrm{bd}$) tested by the 15~$\times$~15 probe card at \SI{20}{\celsius}.}
    \label{USTCIV15x15matrix}
\end{figure}

More than 100 sensors were measured with the 15~$\times$~15 probe card during the prototyping stage and about 10 sensors were measured during the pre-production stage. The measured uniformity of the leakage currents is found to always be improved owing to the simultaneous grounding of all pads and the GR. A typical example is presented here. Compared with the $I-V$ curves of CQC shown in Fig.~\ref{USTCW3P5IVneedles}, it is obvious that the $I-V$ curves tested by the 15~$\times$~15 probe card become more uniform in Fig.~\ref{USTCW3P5IV15x15matrix}. The Max($I$ at 0.8~$\cdot~V_\mathrm{bd}$)/Min($I$ at 0.8~$\cdot~V_\mathrm{bd}$) of the sensor that includes the pads in the outer two rings is less than 3 (see Fig.~\ref{USTCW3P5DicedCR_UBM}). The breakdown voltage of the sensor is about 10 V lower than the results of the CQC measurements, which is expected to be caused by the difference in temperature. The temperature of the chuck was not monitored during the CQC measurements at USTC and the average temperature (approximately \SI{28}{\celsius}) was estimated by remeasuring a sensor with the temperature controlled chuck with the USTC SQC setup. It was found that the CQC measurement can be re-produced with the chuck temperature set to \SI{28}{\celsius}.

\begin{figure}[htbp]
	\centering
        \subfigure[]{\includegraphics[width=0.45\textwidth]{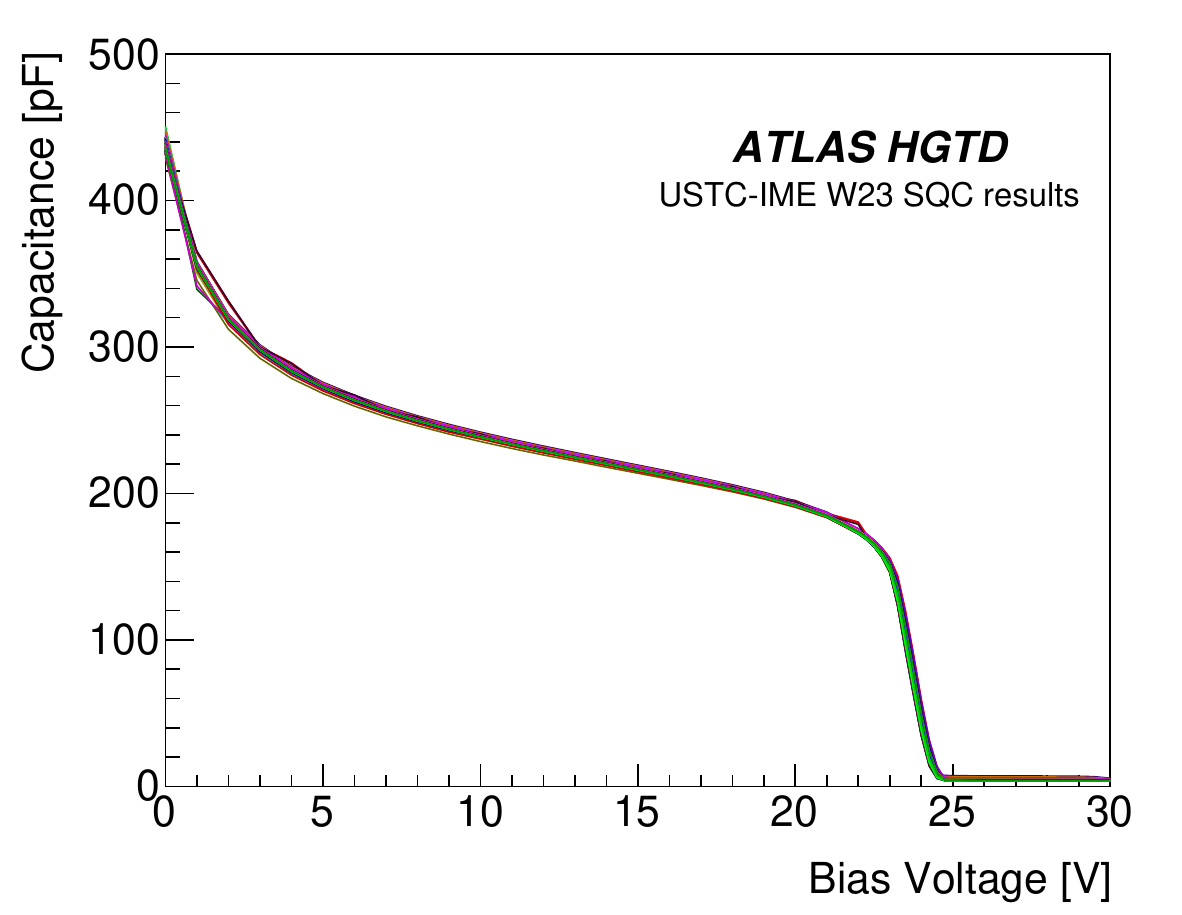}\label{USTCW3P5CV15x15matrix}}
        \subfigure[]{
	    \includegraphics[width=0.45\textwidth]{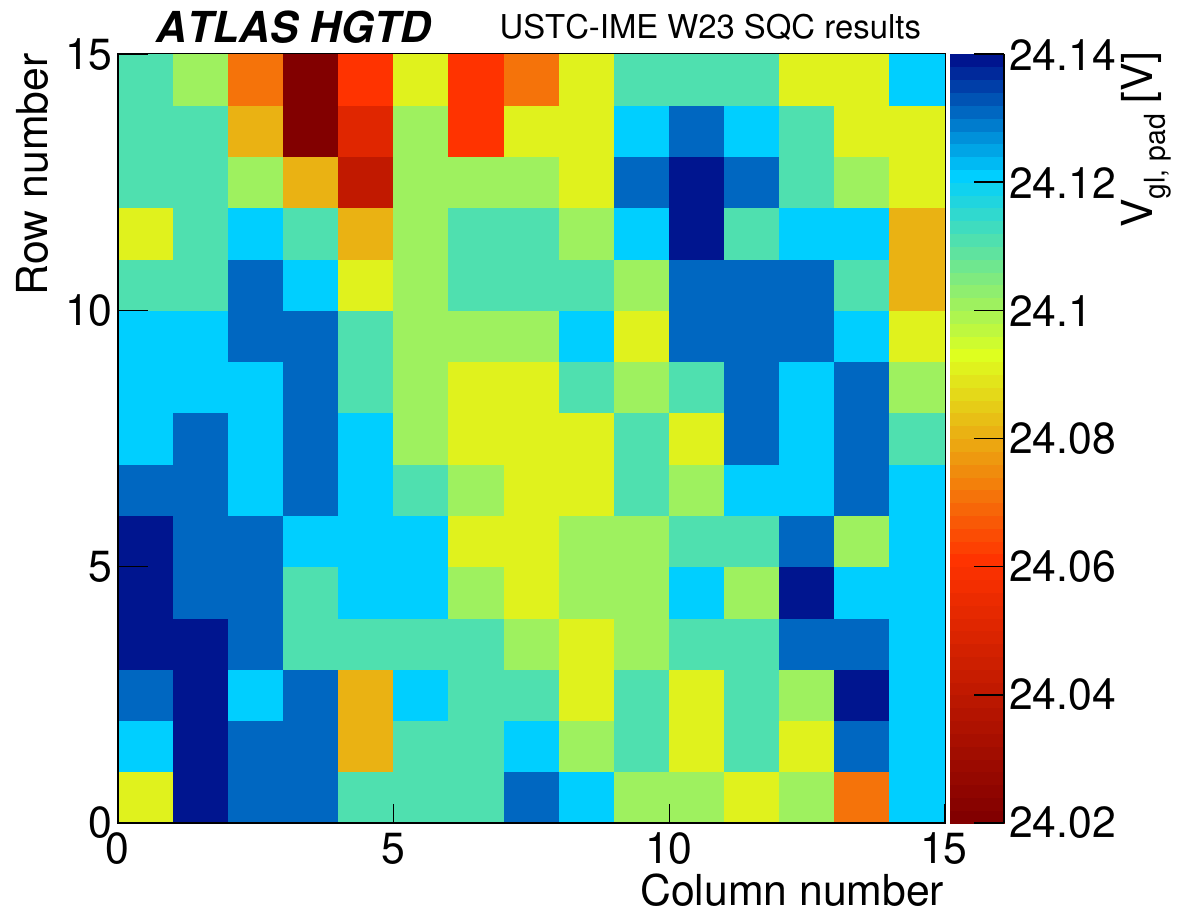}\label{USTCW3P5DicedVGL_UBM}}
        \caption{The $C-V$ results measured by 15$\times$15 probe card. (a) The $C-V$ curves. (b) The 2D distribution of $V\mathrm{_{gl,~pad}}$.}
        \label{USTCCV15x15matrix}
\end{figure}

Based on the $C-V$ curves in Fig.\ref{USTCW3P5CV15x15matrix}, the $V\mathrm{_{gl,~pad}}$ was extracted using the $1/C^2-V$ method~\cite{Ferrero:2021lwf}. In Fig.\ref{USTCW3P5DicedVGL_UBM}, the relative spread extracted from the $V\mathrm{_{gl,~pad}}$ values is about 0.0008, which is much smaller than the requirement (0.005) and also verifies the good uniformity of the sensor.

\section{Irradiation tests}
\label{Irradiation tests}

The Irradiation Test (IT) is a procedure to monitor the sensor radiation hardness. It is carried out on dedicated test structures (QC-TS) with a default frequency of one test per wafer. A method based on the Transient Current Technique (TCT) has been developed to extract four sensor performance parameters: LGAD gain layer depletion voltage, LGAD gain factor at a fixed bias voltage, leakage current, and width of the inactive inter-pad region~\cite{VELKOVSKA2026171195}. To verify the radiation hardness of the gain layer and the isolation structure between pads, an electrical characterization of the irradiated full-size sensors was performed with the 15~$\times$~15 or 5~$\times$~5 probe card, and two parameters were extracted: bias voltage at a fixed leakage current and inter-pad resistance. These quantities will not be used for the SQC during production but was done for the pre-production sensors to check the behavior.

\subsection{Results of $I-V$ test with the 15~\textit{×}~15 probe card}

\begin{figure}[htbp]
	\centering
        \subfigure[]{\includegraphics[width=0.45\textwidth]{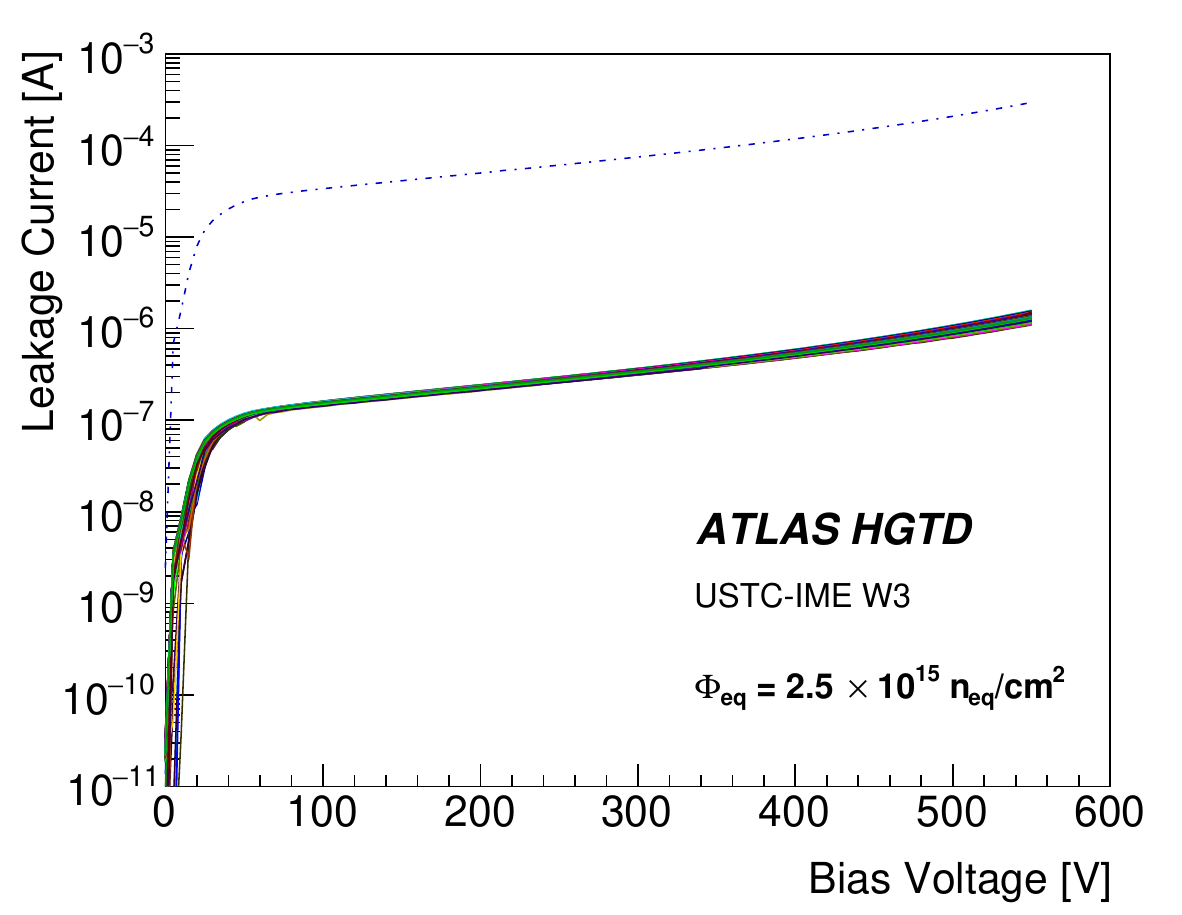}\label{USTCW3P2DicedIV_UBMirr}}
        \subfigure[]{\includegraphics[width=0.45\textwidth]{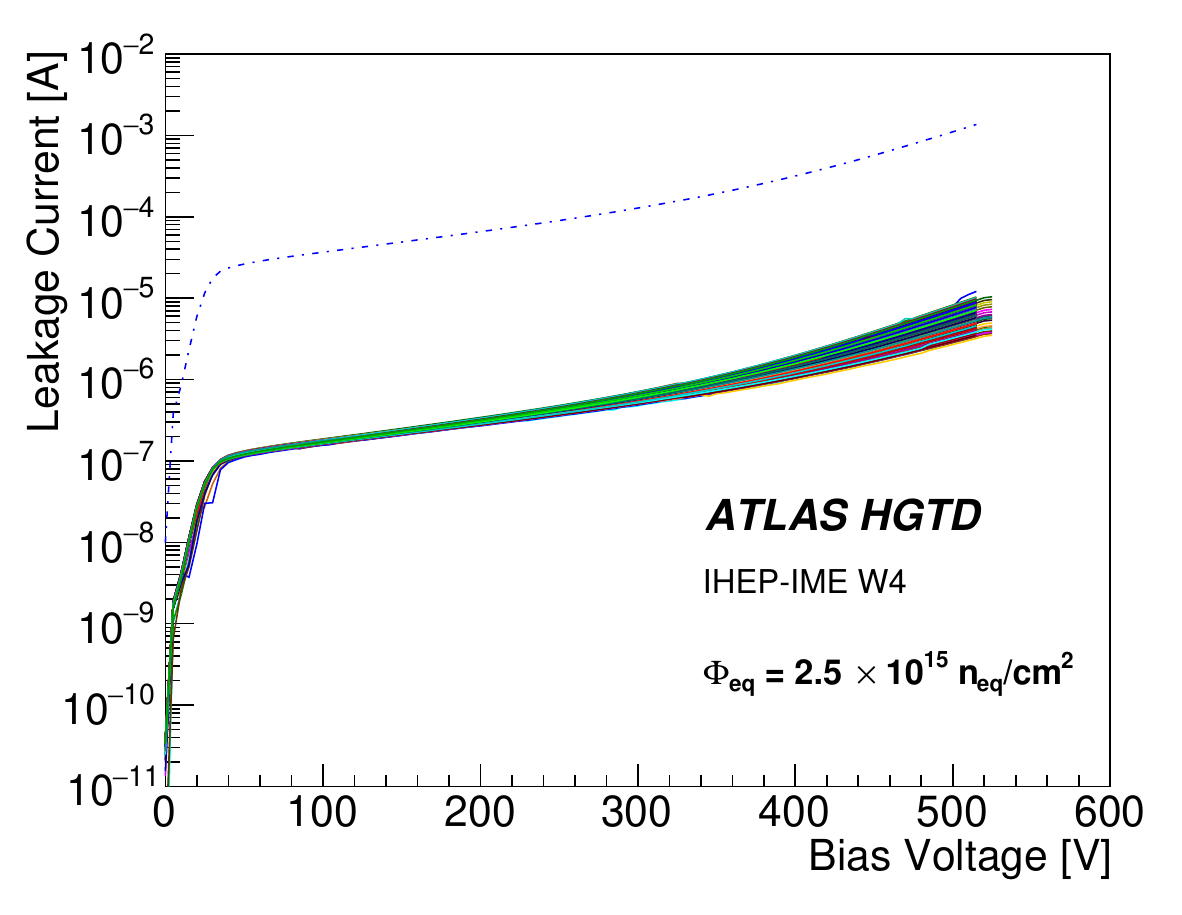}\label{IHEPW4P11DicedIV_UBMirr}}
        \subfigure[]{\includegraphics[width=0.45\textwidth]{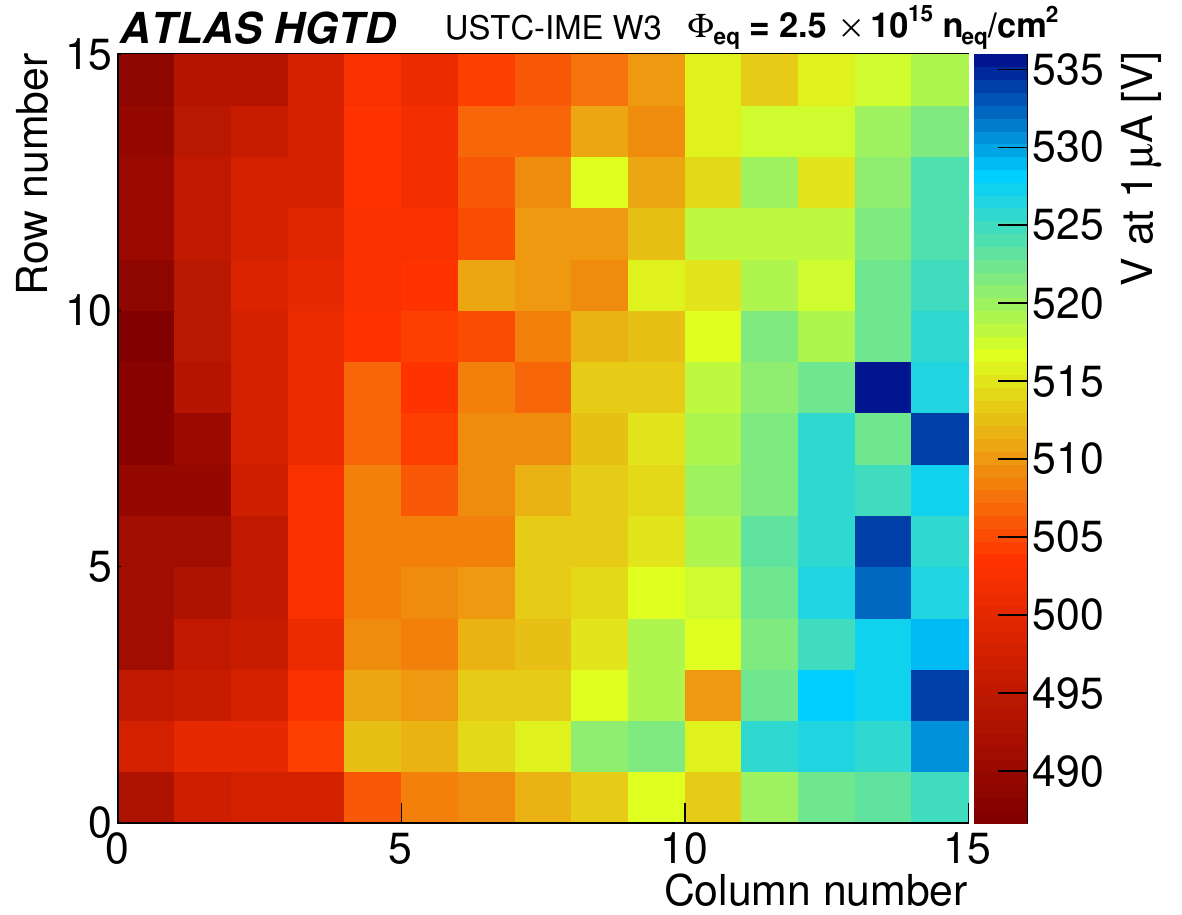}\label{USTCW3P2DicedVBD_UBMirr}}
        \subfigure[]{\includegraphics[width=0.45\textwidth]{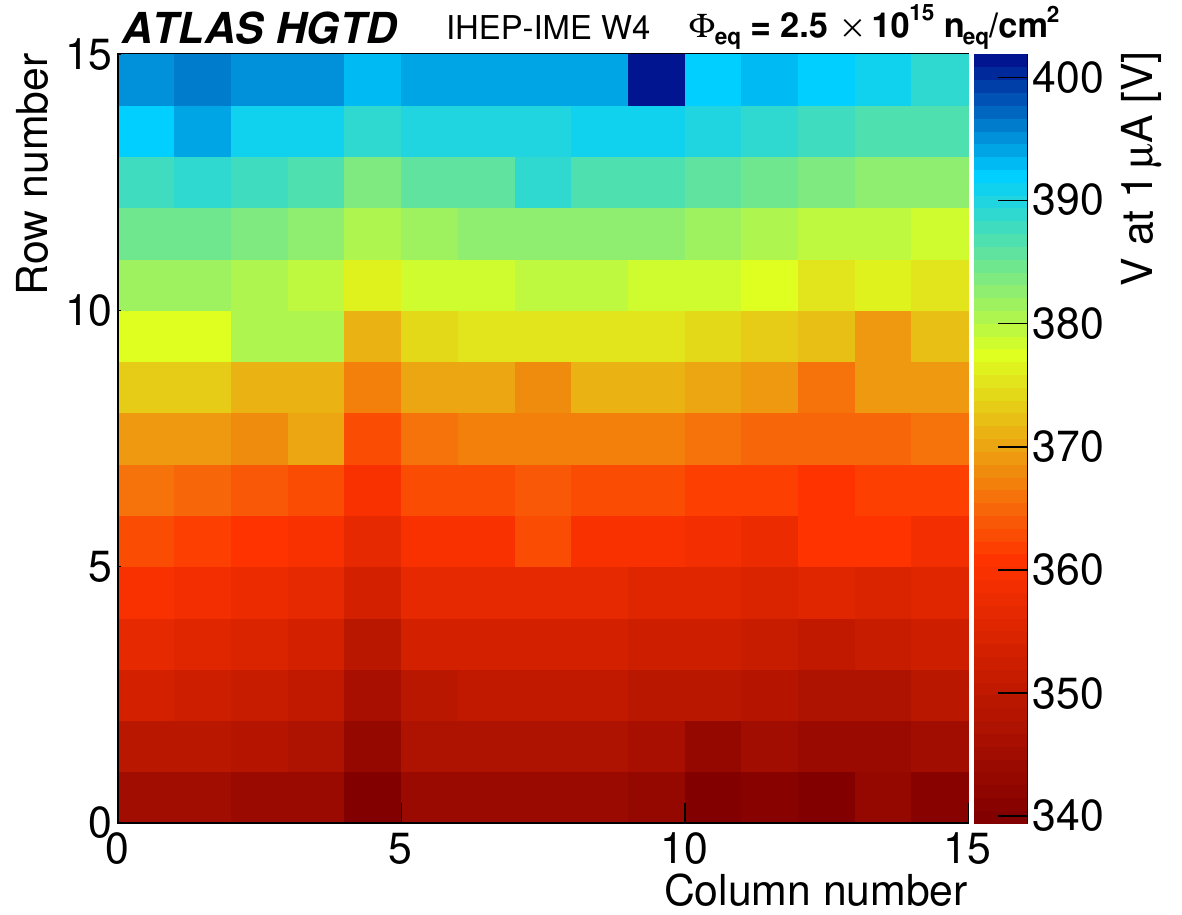}\label{IHEPW4P2DicedVBD_UBMirr}}
        \caption{The $I-V$ results of two typical 15~$\times$~15 array sensors, irradiated by reactor neutrons at JSI with a fluence of 2.5~$\times$~10$^{15}$ n$\mathrm{_{eq}}$/cm$^2$ and measured at \SI{-30}{\celsius}. The $I-V$ curves of a USTC-IME sensor (a) and an IHEP-IME sensor (b). The solid lines represent the pad current and dash-dotted line represents the total current. The 2D distribution of $V$ at \SI{1}{\micro A} of a USTC-IME sensor (c) and an IHEP-IME sensor (d).}
        \label{USTCIV_neutronirr}
\end{figure}

The sensors were irradiated by reactor neutrons at JSI~\cite{Snoj:2012dib} with a fluence of 2.5~$\times$~10$^{15}$ n$\mathrm{_{eq}}$/cm$^2$ and annealed at \SI{60}{\celsius} for 80 minutes before testing. Fig.~\ref{USTCIV_neutronirr} shows the $I-V$ results of two typical irradiated 15~$\times$~15 array sensors tested at \SI{-30}{\celsius}. As shown in Figs.~\ref{USTCW3P2DicedVBD_UBMirr} and ~\ref{IHEPW4P2DicedVBD_UBMirr}, the $V$ at \SI{1}{\micro A} of all pads for both USTC-IME and IHEP-IME sensors falls below 550 V, indicating that these irradiated sensors can achieve sufficient charge before SEB. However, the 2D distributions of $V$ at \SI{1}{\micro A} exhibit a large spread, which is attributed to the non-uniformity of the irradiation fluence across the large area of the sensor (approximately 20 mm~$\times$~20 mm). 

\subsection{Inter-pad resistance}

An inter-pad resistance measurement was performed to obtain information on the quality of p-stop isolation implanted between the pads. Surface damage leads to a decrease of the inter-pad resistance~\cite{Hartmann:2017gzy}. The charging up of the oxide with positive charge forms a conductive layer between two n$^+$ structures. Therefore, a p-stop layer is introduced to isolate the pads. The p-stop isolation can be degraded in case of production issues or for a too large radiation-induced accumulation of positive charges. 

Different types of damage have been investigated through irradiations at the different facilities. In addition to the sensors irradiated by reactor neutrons at JSI, several sensors were irradiated by $^{60}$Co gamma rays at RBI Zagreb~\cite{MAJER2024111280} to a TID dose of 1.5 MGy, which is close to the dose expected for the HGTD at the HL‑LHC. 

\begin{figure}[htbp]
	\centering
        \subfigure[]{\includegraphics[width=0.45\textwidth]{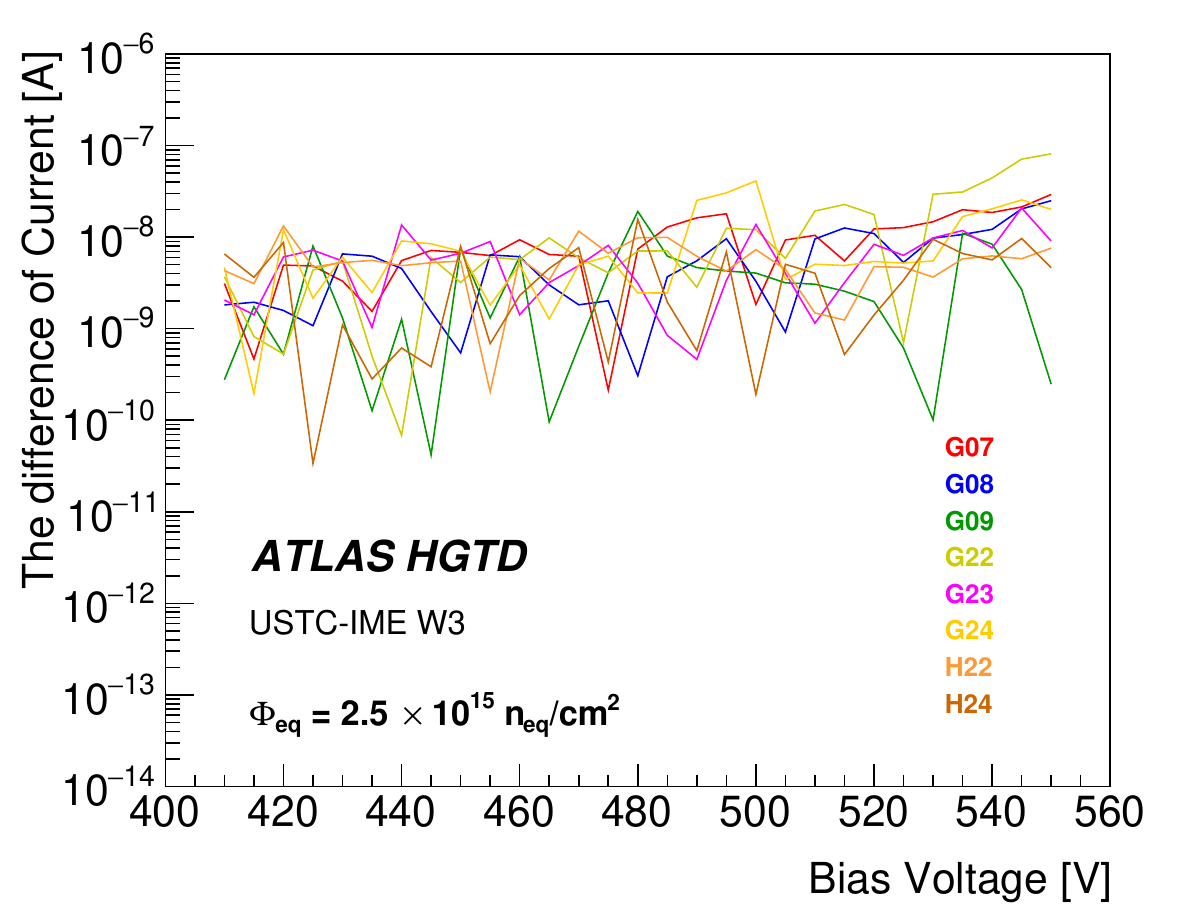}\label{USTCW3P2IPneutron_UBMirr}}
        \subfigure[]{\includegraphics[width=0.45\textwidth]{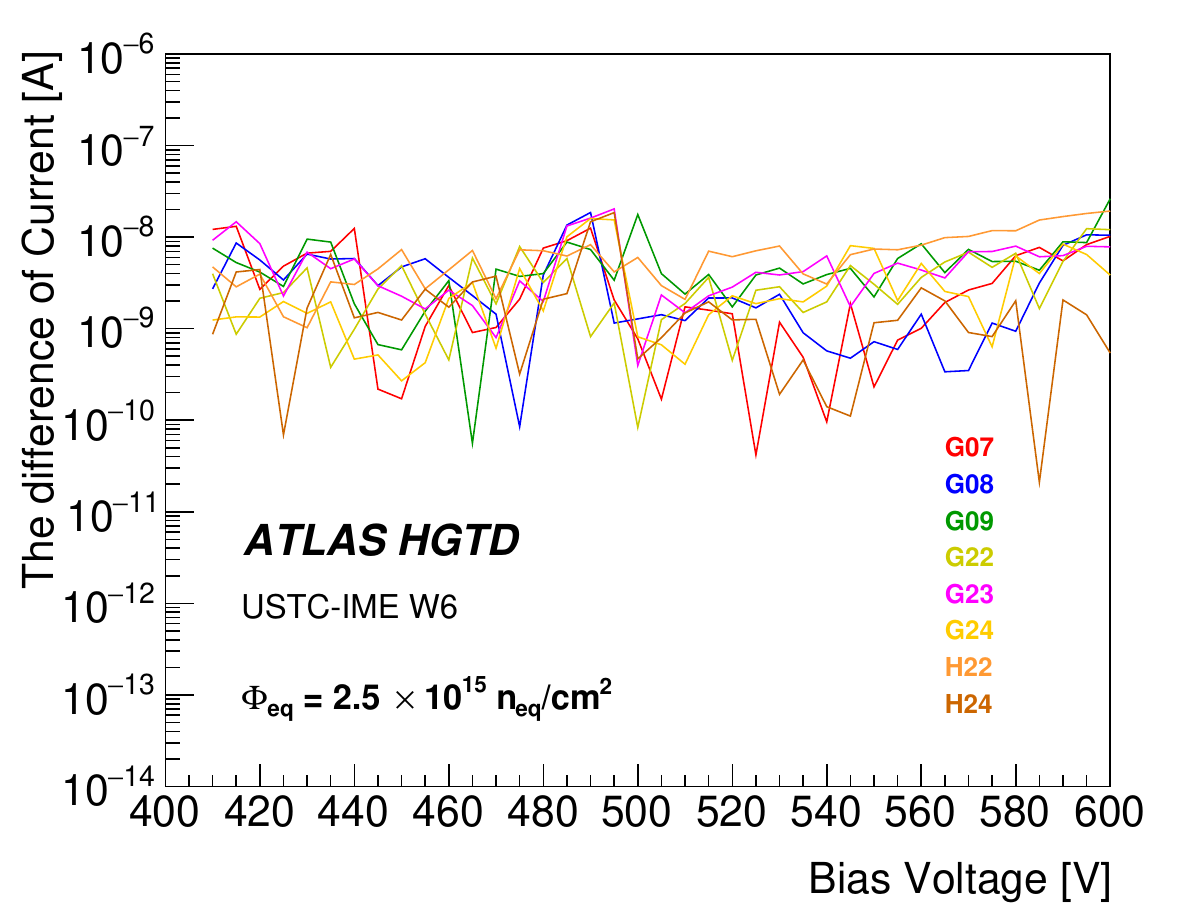}\label{USTCW6P3IPneutron_UBMirr}}
        \caption{Inter-pad resistance measurements of the irradiated 15~$\times$~15 array sensors at \SI{-30}{\celsius}. Samples were irradiated by reactor neutrons at JSI with a irradiation fluence of 2.5~$\times$~10$^{15}$ n$\mathrm{_{eq}}$/cm$^2$. The difference of current as a function of bias voltage, which is caused by the difference of 1 V on the central pad: (a) a USTC-IME sensor on wafer 3 and (b) a USTC-IME wafer on wafer 6.}
        \label{USTCIP_neutronirr}
\end{figure}

Figs.~\ref{USTCW3P2IPneutron_UBMirr} and ~\ref{USTCW6P3IPneutron_UBMirr} show the difference of current when the bias voltage is changed by 1 V, of two sensors  irradiated by neutrons with a fluence of 2.5~$\times$~10$^{15}$ n$\mathrm{_{eq}}$/cm$^2$. As the bias voltage increases, the difference of current is always less than 100 nA, indicating that the inter-pad resistance is greater than \SI{10}{\mega\ohm}. 

\begin{figure}[htbp]
	\centering
        \subfigure[]{\includegraphics[width=0.45\textwidth]{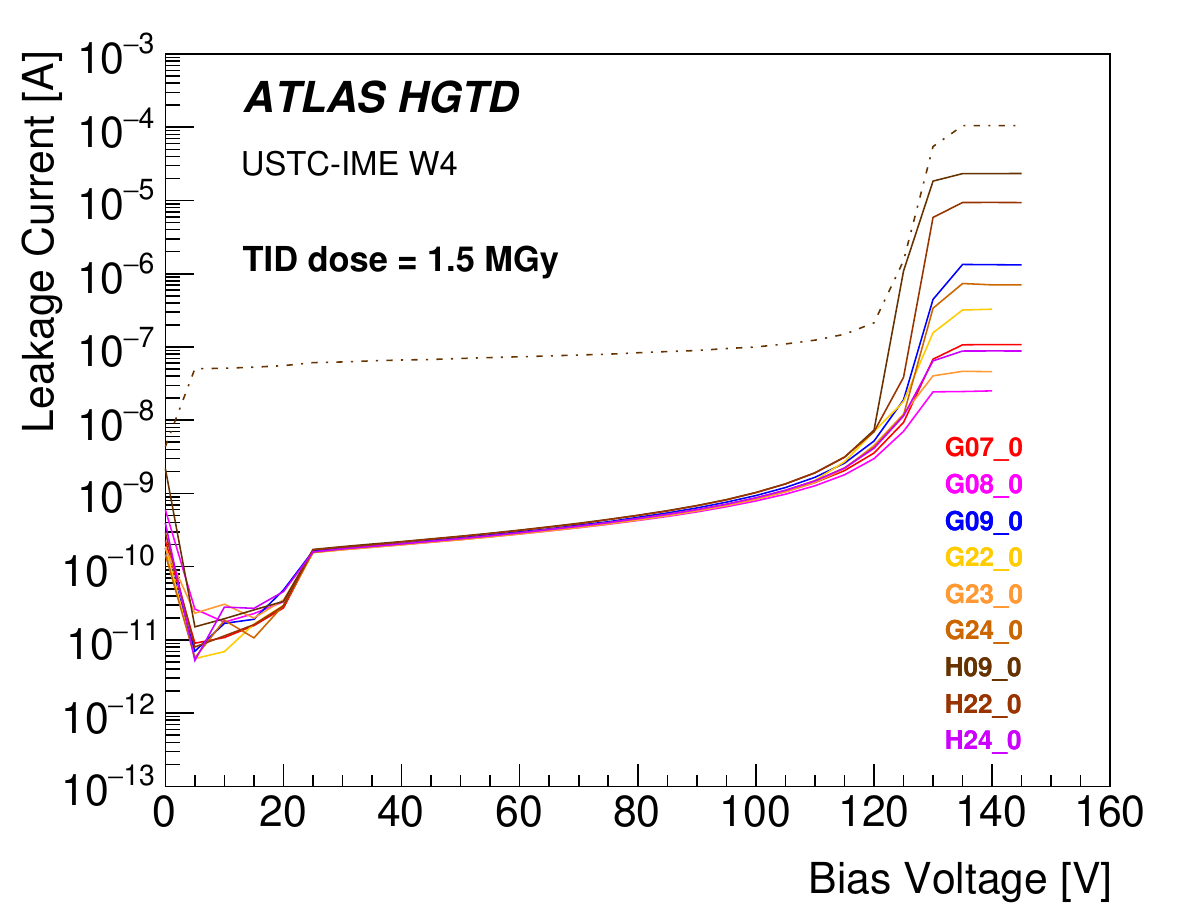}\label{USTCW4P13DicedIV_UBMprotonirr}}
        \subfigure[]{\includegraphics[width=0.45\textwidth]{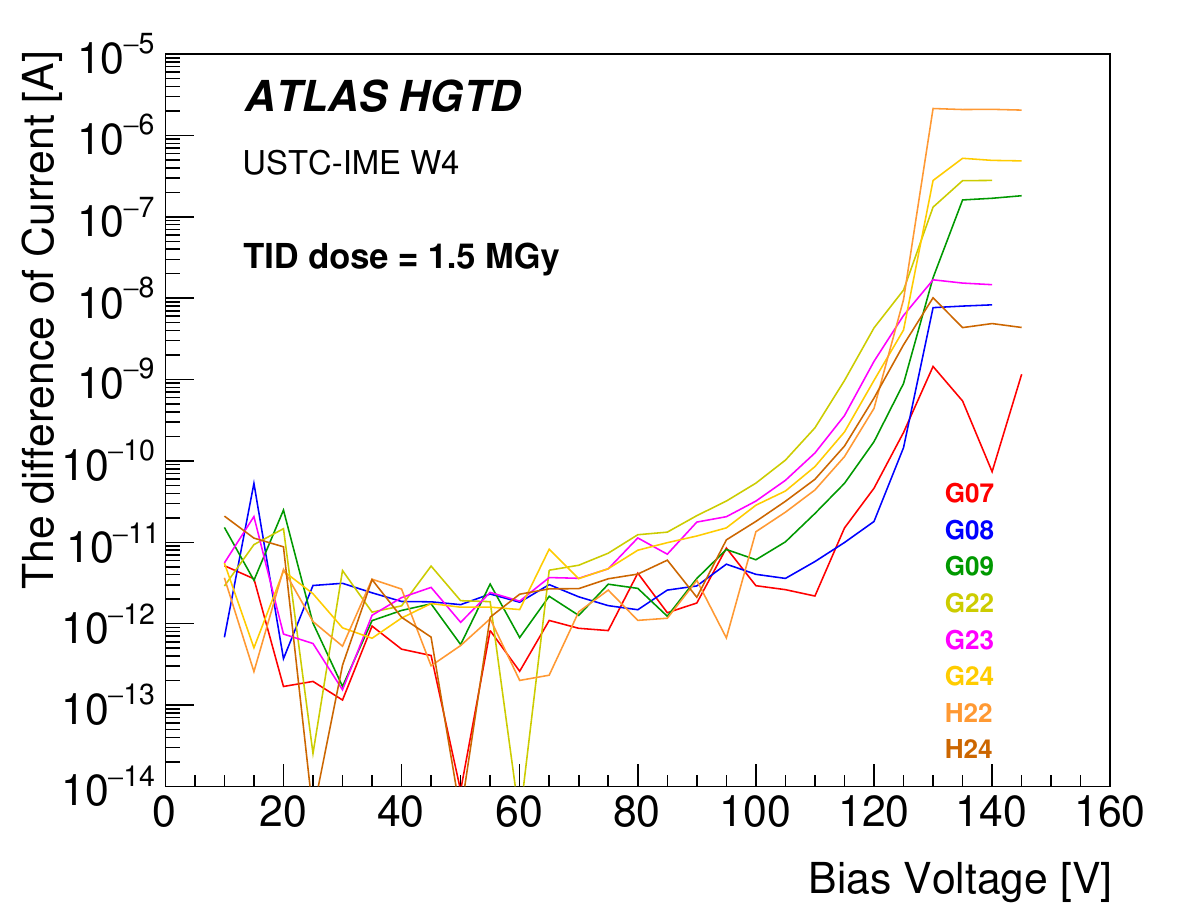}\label{USTCW4P13IPproton_UBMirr}}
        \caption{Inter-pad resistance measurements of the irradiated 15~$\times$~15 sensors at \SI{-30}{\celsius}. The sample was irradiated by $^{60}$Co gamma with an irradiation dose of 1.5 MGy. (a) The $I-V$ curves of the eight adjacent pads (G07, G08, G09, G22, G23, G24, H22 and H24) and the total current of 25 pads (dash dotted line) tested by 5~$\times$~5 probe card when applying 0~V on the central pad (H09). (b) The difference of current as a function of bias voltage, which is caused by the difference of 1 V on the central pad.}
        \label{USTCIP_protonirr}
\end{figure}

Fig.~\ref{USTCIP_protonirr} shows the results of a sensor irradiated by $^{60}$Co gamma rays with a dose of 1.5 MGy. Fig.~\ref{USTCW4P13DicedIV_UBMprotonirr} shows the $I-V$ curves of the eight adjacent pads (G07, G08, G09, G22, G23, G24, H22 and H24) and the total current of 25 pads (dash-dotted line) tested by the 5~$\times$~5 probe card when applying 0 V on the central pad (H09). Due to the total-current compliance setting of \SI{10}{\micro A}, the individual leakage currents reach a plateau below this value for the last bias points. The difference of current in Fig~\ref{USTCW4P13IPproton_UBMirr} is at a very low level at low bias voltage; then it increases with the bias voltage and it can be greater than 100 nA when these 25 pads break down. The inter-pad resistance is larger than \SI{10}{\mega\ohm} at the maximum operation voltage of 0.8~$\cdot~V\mathrm{_{bd}}$. For reference, the inter-pad resistance of unirradiated sensors was measured with the same method to be around \SI{100}{\giga\ohm}. As expected, the isolation of the structures is degraded with irradiation, but the resistance remains within the specifications up to the expected end-of-life doses and fluences. 

Since the oxide growth was fabricated using the same production line equipment and both the USTC-IME and IHEP-IME sensors employ the p‑stop isolation structure, it is reasonable to assume that the inter‑pad resistance of the IHEP‑IME sensors will also meet the requirements.

\section{Conclusion}
\label{Conclusion}
A fast and reliable quality control is required for the ATLAS-HGTD sensor production to select full-size LGAD sensors, where the requirements for breakdown voltage and leakage current vary according to the sensor design and test configuration. CQC based on $I-V$ measurements using the per-pad test method or 15~$\times$~1 test method, is introduced to extract breakdown voltage and leakage current on the wafer level. Applying a stricter criterion on leakage current uniformity for the 15~$\times$~1 test method yields results consistent with per-pad testing. The per-pad test results of CQC have been compared with SQC, which includes the per-pad test and 15~$\times$~15 matrix test, and the reliability of the CQC has been demonstrated. The 15~$\times$~15 matrix tests, which is consistent with the actual operating state of the sensor (except for floating of GR), indicate that the spread of current and $V\mathrm{_{gl,~pad}}$ is more uniform under these operating conditions. Moreover, tests of irradiated sensors based on $I-V$ measurements are performed, showing radiation tolerance of the gain layer and good isolation of the inter-pad region after receiving radiation damage up to 2.5~$\times$~10$^{15}$ n$\mathrm{_{eq}}$/cm$^2$ or 1.5 MGy. In general, good agreement was found between the CQC and the detailed SQC measurements. The per-pad $I-V$ measurement and the $15~\times~1$ probe card measurement will be used for CQC during the production phase. The corresponding acceptance criteria have been refined and validated in this work.

\acknowledgments

This work was supported by the National Natural Science Foundation of China (No. 11961141014), the Ministry of Science and Technology of China (No. 2023YFA1605901), China; the Slovenian Research and Innovation Agency (ARIS P1-0135, ARIS Z1-50011, ARIS J7-4419), Slovenia; and FAPESP (2020/04867-2, 2022/14150-3, 2023/18486-9 and 2023/18484-6), MCTI/CNPq (INCT CERN Brasil 406672/2022-9) and CAPES - Code 001, Brazil. We are grateful for staff members involved in this work at IME.





\bibliographystyle{JHEP}
\bibliography{biblio.bib}

\end{document}